# Rational Designing of Anthocyanidins-Directed Near-Infrared Two-Photon Fluorescence Probes

Xiu-e Zhang[a], Xue Wei[b], Wei-Bo Cui[b], Jin-Pu Bai[a], Aynur Matyusup[a], Jing-Fu Guo*[a], Hui Li[b] and Ai-Min Ren*[b]

[a]School of Physics, Northeast Normal University, Changchun 130024, P.R.China

[b]Laboratory of Theoretical and Computational Chemistry, Institute of Theoretical Chemistry, College of Chemistry, Jilin University, Liutiao Road #2, Changchun 130061, P.R.China

Corresponding Author: Jing-Fu Guo

E-mail: guojf217@nenu.edu.cn.

ORCID: 0000-0002-1864-167X

Corresponding Author: Ai-Min Ren

E-mail: renam@jlu.edu.cn.

ORCID: 0000-0002-9192-1483

# Abstract

Recently, two-photon fluorescent probes based on anthocyanidins molecules have attracted extensive attention due to their outstanding photophysical properties. However, there are only a few two-photon excited fluorescent probes that really meet the requirements of relatively long emission wavelengths (>600 nm), large two-photon absorption (TPA) cross sections (300 GM), significant Stokes shift (>80 nm), and high fluorescence intensity. Herein, the photophysical properties of a series of anthocyanidins with the same substituents but different fluorophore skeletons were investigated in detail. Compared with **b**-series molecules, **a**-series molecules with a six-membered ring in the backbone have a slightly higher reorganization energy.. This results in more energy loss upon light excitation, enabling the reaction products to detect NTR through a larger Stokes shift. More importantly, there is very little decrease in fluorescence intensity as the Stokes shift increases. These features are extremely valuable for high-resolution NTR detection. In light of this, novel **2a-n** (**n**=**1**-**5**) compounds are designed, which are accomplished by inhibiting the twisted intramolecular charge transfer (TICT) effect through alkyl cyclization, azetidine ring and extending π conjugation. Among them, **2a-3** gains long emission spectrum ($\lambda_{em}$=691.42 nm), noticeable TPA cross section (957.36 GM), and large Stokes shift (110.88 nm), indicating that it serves as a promising candidate for two-photon fluorescent dyes. It is hoped that this work will offer some insightful theoretical direction for the development of novel high performance anthocyanin fluorescent materials.

# 1. Introduction

Nitroreductase (NTR) in hypoxic tumors would likely be overexpressed, making it critical to precisely and in real-time monitor NTR.[1–3] Small molecule fluorescent imaging technology has widespread adoption in diverse fields in recent years due to its numerous merits, such as superior biocompatibility, rapid and sensitive feedback, non-invasive, and easy adjustment of optical properties through structural modification.[4–6] Unlike the conventional one-photon (OP) excitation probes, the two-photon (TP) fluorescent probes have received considerable attention because of their excitation wavelengths in the near-infrared (NIR) region.[7] This characteristic provides various unique advantages, including negligible background fluorescence interference, minimal photodamage to biological samples, substantially decreased photobleaching, and deeper tissue penetration.[8–12] Above all, two-photon excited fluorescence (TPEF) probes have emerged as powerful tools in both fundamental biological research and clinical applications.[13–16]

Although a number of molecular fluorophores have been reported in the past few decades, the relatively short wavelength (<600 nm) still constrains their applications in the biological fields.[17–20] In this case, our attention is piqued by the fact that anthocyanidin, a natural pigment, can achieve long wavelength emission with great selectivity and sensitivity.[21,22] A unique fluorescent molecular framework called AC-Fluor was designed in 2017 and its biological relevance was proved using two-photon deep tissue imaging, exhibiting deep penetration up to 300 μm with negligible cytotoxicity.[23] In 2019, their group synthesized several kinds of anthocyanidins based on **1b** (molecule **ACF4** in the literature) in **Fig. 1(a)**.[24] In particular, **LDO-NTR** was proposed as a practical NTR-activated TPEF probe as a result of molecule **2a**, also known as **LDOH-4** in the literature, demonstrating favorable chemical and optical properties ($\lambda_{em}$=633 nm, $\Phi$=0.55, p$K_a$=5.13, Stokes shift=59 nm). Currently, there is no comprehensive and in-depth investigation about anthocyanidin derivatives and their unique electronic structures as TPEF fluorescence probes, which has caught our considerable attention and enthusiasm. Beyond that, it appears that the photoinduced electron transfer (PET) and intramolecular charge transfer (ICT) results in the fluorescence quenching of **LDO-NTR**. However, the underlying microscopic mechanism of fluorescence quenching still remains ambiguous. Besides, for **LDO-NTR** derivatives with the same substituents but different fluorophore skeletons **1a-3a** (corresponding to **LDOH-1**, **LDOH-4**, **Ctrl-2** in the reference[24]) and **1b-3b** (corresponding to **ACF4**, **LDOH-3**, **Ctrl-1** in the reference[24]) show excellent properties such as suitable p*K*a, high quantum yield, and good photostability. How do the six/five-membered rings in the backbone affect their photophysical properties differently? Regarding molecule **2a**, also, is there any room to improve its optical properties, such as small two-photon absorption (TPA) cross section (89 GM/$\Phi$=161.8 GM), short emission wavelength (633 nm) and small Stokes shift (59 nm)? If so, further application opportunities would arise as it would improve the photobleaching limit and fluorescence/background ratio of anthocyanin-based TP fluorescent probes for super-resolution and deep tissue imaging, opening up broader application prospects. All of the issues aforementioned are worthy of serious attention and need to be addressed urgently.

In this work, density functional theory (DFT) and time-dependent density functional theory (TDDFT) are adopted to systematically reveal the relationship between molecular configurations and properties of a series of anthocyanidin derivatives (molecular probes with 4-nitrobenzyl alcohol group as the NTR reaction site and the corresponding products with hydroxyl group), including OPA/TPA properties, fluorescence emission properties, fluorescent probing mechanism, and solvation free energy. Then, new series of **2a-n** (**n**=**1**-**5**) TPEF molecules are designed with accounting for the biological applications of TP fluorescent probes. Ultimately, the PET and ICT luminescence mechanism of the anthocyanidin derivatives are investigated by means of quantum chemistry methods. These findings will hopefully provide some meaningful insights into the rational design of novel functional fluorescent materials based on the anthocyanidin skeleton.

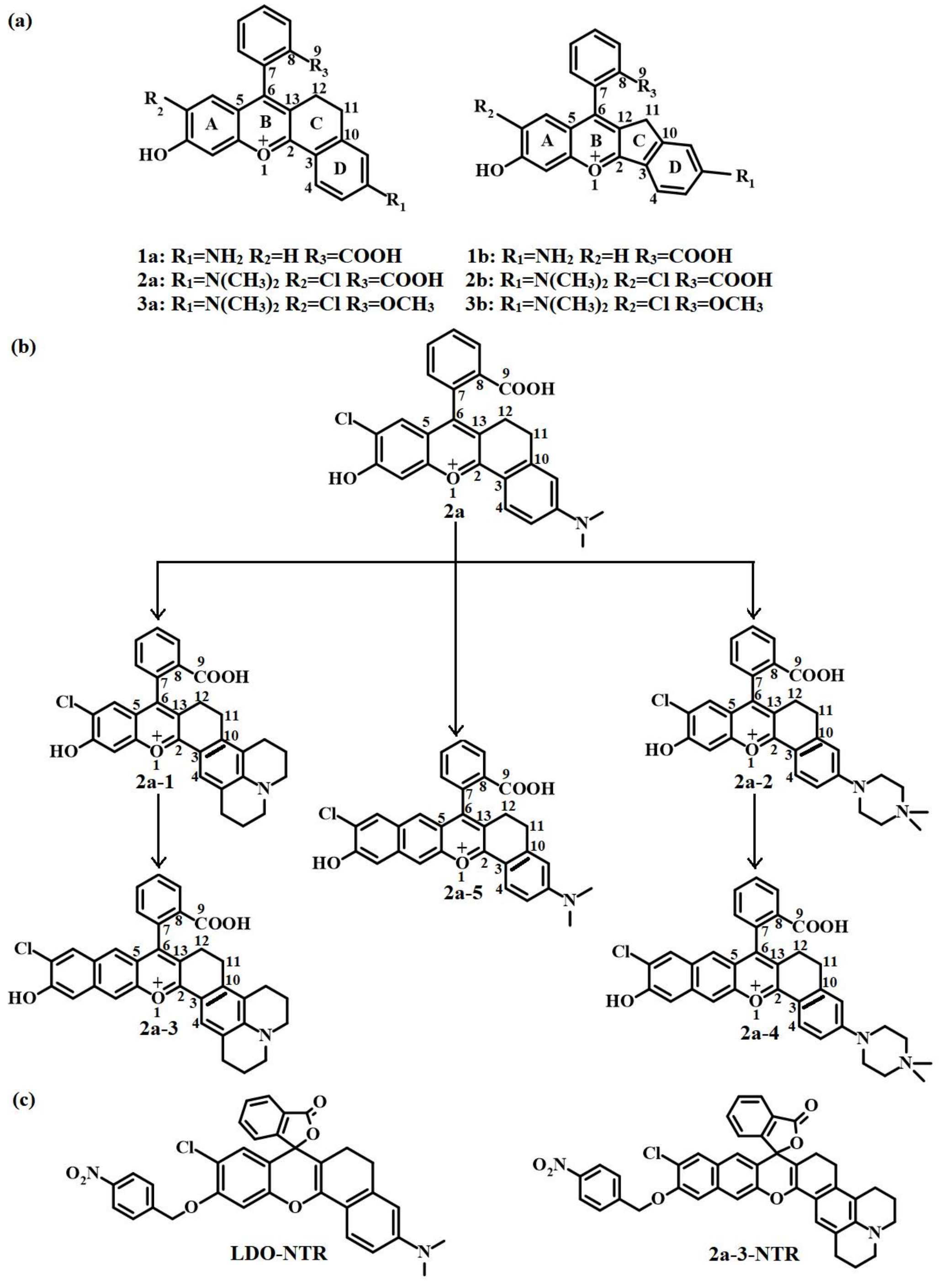

**Fig. 1** The molecular structures and atomic numbering of (a) the studied experimental molecules (b) the designed **2a-n** (**n=1-5**) TPEF molecules (c) two typical probe molecules.

# 2. Calculation Methods

The full explanation of theoretical methodologies utilized in this paper are given in **Supporting Information**, including one-photon absorption (OPA), two-photon absorption (TPA), and solvation free energy. In addition, the computational details are

summarized as follows:

The geometric structures of the ground states and excited states of the studied molecules were optimized using DFT[25] and TDDFT,[26] respectively. Vibrational frequency analysis was performed based on the optimized geometric configurations, and no imaginary frequency was emerged. All of the preceding above calculations were carried out in Gaussian 16 program. Taking the experimental molecules **1a** and **1b** as references, their OPA and fluorescent emission spectra were calculated by TDDFT utilizing different functionals and basis sets. Detailed results are listed in **Table S1-S3**. As a compromise between accuracy and computational efficiency, B3LYP/6-31G(d, p) and B3LYP/6-311+G(d) methods are chosen and employed in the subsequent optimization and transition energy calculation, respectively. On the basis of quadratic response theory,[27] the B3LYP/6-311+G(d) and CAM-B3LYP/6-311+G(d) method were also applied to calculate the TPA properties of the experimental molecule **2a** as an example using the DALTON[28] program. As can be seen in **Table S4**, the result for B3LYP (185.73GM) is closer to the experimental value (161.8 GM) than CAM-B3LYP (581.00GM). Therefore, the TPA properties of all the subsequent molecules were calculated using B3LYP/6-311+G(d). Furthermore, a polarization continuous model (PCM)[29] was used to take into account the solvent effect. Previous efforts have confirmed that the M052X/6-31G(d) together with the SMD solvent model is an optimal approach for solvation free energy calculations,[30] thus the method was applied in this work. The superimposed structures of $S_0$ and $S_1$ states were achieved by PyMOL program.[31]

# 3. Results and Discussion

## 3.1. Molecular Design

Firstly, a series of anthocyanidins with the same substituents but different fluorophore skeletons, **1a-3a** and **1b-3b**, are selected in an effort to study how do the six/five-membered rings in the backbone affect their photophysical properties. The chemical structures of the studied experimental compounds are presented in **Fig. 1(a)**. The maximum absorption peak (503.80~535.52 nm) of **b**-series anthocyanidins with five-membered ring fused in the backbone have a hypsochromic shift with respect to **a**-series with six-membered ring fused in the backbone (519.95~551.59 nm). Moreover, the **b**-series molecules exhibit stronger fluorescence oscillator intensity (1.1269~1.2102) than that of **a**-series (1.0430~1.1251). What role does the five and six-member rings in the backbone play on the photophysical property? In an attempt to address the above issues, **1c** molecule is also designed as a reference, and its chemical structure is shown in **Fig. S1**. Their electronic structures and photophysical properties, such as OPA and TPA spectra, are scrutinized in the sections **3.2-3.5**.

Secondly, it is worth highlighting that **2a** exhibits superior properties for biological imaging, with a better fluorescence enhancement factor (27.8-fold) for phenol/phenolate states than **2b** (23.7-fold), and its excellent properties under simulated physiological conditions (**2a**: $\lambda_{em}$ = 633 nm, $\Phi$ = 0.55, pKa = 5.13; **2b**: $\lambda_{em}$ = 620 nm, $\Phi$ = 0.70, p*K*a = 5.60)[24]. In this context, some design strategies are put forward by us based on the experimentally synthesized molecule **2a** in the hope of improving properties such as emission wavelength and TPA cross sections. In addition, the molecules **4a** and **5a** (as shown in **Fig. S7**) are intended to function as a comparison for **2a** to identify the effects from the various substitution sites (the detailed discussion are provided in **Supporting Information**). Eventually, the five compounds (**2a-n** (**n**=**1-5**)) are designed by modifying $R_1$ substituent group on the basis of the molecule **2a** with a six-membered ring that is not fully conjugated, and **Fig. 1(b)** offers a reference to their unique molecular structures. Most notably, dialkylamino groups, which are extensively utilized as electron-donating groups in traditional fluorophores, are regarded as the rotating groups resulting in twisted intramolecular charge transfer (TICT).[32] The non-radiative decays of some fluorophores coincide with the formation of TICT states, hence it may be quencher of many highly effective fluororescence.[33–35] It has been reported that performing alkyl cyclization and azetidine substitution can effectively suppress TICT formation.[34–38] thereby, **2a-1** and **2a-2** are designed for making every effort to improve the emission properties. To achieve a larger TPA response, **2a-3**, **2a-4** and **2a-5** are also designed by extending the fused backbone. Along with the first part of the work, the research on the photophysical properties of the designed molecules is also included in the sections **3.2-3.5**.

Thirdly, the fluorescence quenching of **LDO-NTR** upon light excitation is regarded to the combined effects from both photoinduced electron transfer (PET) and intramolecular charge transfer (ICT). However, the underlying mechanism of fluorescence quenching phenomenon of the probe still remains ambiguous in fact. We have carried out a detailed theoretical study about the ICT and PET processes of **LDO-NTR** to rationalize the phenomenon. Besides, as illustrated in **Fig. 1(c)**, for designed molecules **2a-n** (**n**=**1**-**5**) series, taking the probe **2a-3-NTR** as an example, its corresponding product to detect NTR reaction, **2a-3**, has the mostly excellent properties among the designed compounds **2a-n** (**n**=**1**-**5**), and its fluorescence quenching mechanism is also predicted by using the same theoretical method. This part is described in the section **3.6**.

## 3.2. Geometrical Optimization

Geometric configurations are intimately tied to electronic structures and further photophysical properties. Therefore, some selected geometric parameters of the studied molecules at the optimized $S_0$ and $S_1$ states are summarized in **Table S5**, and their superimposed structures are depicted in **Fig. S2**. Upon conducting a comparative analysis, it can be concluded that **a**-series of molecules relative to **b**-series with the same substituents but different fluorophore skeletons, (1) possess longer interatomic distance ($C_2$-$C_3$, $C_{10}$-$C_{13}$) (The numbering of atoms are shown in **Fig. 1**), indicating that **a**-series molecules might enhance the molecular structure stability by weakening the interaction between the **B** and **D** rings.[39] (2) the dihedral angle (DHA1: 6-13-12-11 is 147~150 degree for **a**-series of molecules or 6-12-11-10 are 177~178 degree for **b**-series molecules) of **a**-series molecules is not as close to 180 degree as that of **b**-series molecules, which demonstrates that the former exhibit less planarity. (3) have the least root-mean-square deviation (RMSD) values, suggesting that **a**-series molecules would experience smaller structural distortion upon excitation. Compared to **2a**, the differences in geometric parameters at $S_0$ states are extremely minor for the designed molecules (**2a-1~2a-5**). However, the RMSD values of **2a-3** and **2a-5** are close to zero, which implies that there is almost no structural distortion during the transition from the $S_0$ to $S_1$ states. (4) from the DHA2 in **Table S5**, the carboxyl-phenyl fragment of **a**-series molecules twists relatively less during the $S_0 \rightarrow S_1$ process than that of **b**-series molecules. Overall, the six-membered ring in **a**-series molecules minimizes the geometry rotation upon light excitation with respect to **b**-series molecules. Nevertheless, as shown in **Table S6**, the reorganization energy of **a**-series molecules (**1a**: 1262.04 $cm^{-1}$, **2a**: 1370.09 $cm^{-1}$, **3a**: 1101.87 $cm^{-1}$) is greater than that of **b**-series molecules (**1b**: 1187.46 $cm^{-1}$, **2b**: 1338.33 $cm^{-1}$, **3b**: 1089.32 $cm^{-1}$). It is surprised that the six-membered ring in **a**-series molecules makes the reorganization energy increase during the $S_1 \rightarrow S_0$ process. This will be discussed in more detail in the following section, leading to an increase in the Stokes shift.

## 3.3. Frontier Molecular Orbitals

To further investigate the properties of electronic structures, molecular orbitals were calculated by DFT//B3LYP/6-31G(d,p). The 10 frontier molecular orbitals (FMOs) of the studied molecules are depicted in **Fig. 2**, in which **Fig. 2(a)** shows that compared to **b**-series with the same substituents but different fluorophore skeletons, **a**-series molecules have a lower LUMO energy level and obviously smaller energy gaps ($\Delta E_{\mathrm{HOMO-LUMO}}$) between HOMO and LUMO. The above phenomenon indicates that the introduction of a six-membered ring in the fluorophore skeleton makes **a**-series molecules have lower LUMO levels than **b**-series molecules. As shown in **Fig. S3**, it is observed that the HOMO and LUMO of the **a**-series and **b**-series molecules have extremely similar electronic cloud distributions. In order to scrutinize the different role of the added five and six-membered ring in backbone, the **1a**, **1b**, and **1c** (unsubstituted on backbone) are taken as examples, and the difference in their LUMO are clarified as follows:

On the one hand, as can be seen from **Fig. S4**, the LUMO distribution of **1a** and **1b** is basically the same, primarily composed of antibondings and nonbondings, with the most intense antibonding located between the $O_1$-$C_2$ bond. **Table S5** indicates that **1b** has a shorter $O_1$-$C_2$ bond length (1.3297 Å) than both **1a** (1.3422 Å) and **1c** (1.3465 Å). This difference means that **1b** has stronger antibonding in LUMO, which ultimately results in a higher LUMO energy level. On the other hand, the same reason mentioned above also explains the wider HOMO-LUMO gap of the **b**-series molecules, the larger HOMO-LUMO gap also further implies that they might be responsible for a hyperchromic OPA spectrum.

Similarly, the FMOs analysis for the designed molecules **2a-n** (**n**=**1-5**) is also performed. As illustrated in **Fig. 2(b)**, it's clear that compared with the experimental molecule **2a** (2.65 eV), the $\Delta E_{\text{HOMO-LUMO}}$ of **2a-1** (2.54 eV) decreases dramatically, while **2a-2** (2.76 eV) exhibits the opposite behavior. This is due to the fact that **2a-1** shows increased the contribution of alkyl cyclization substituent in HOMO relative to the **2a** (shown in **Fig. S11**), and the substituent is not directly involved in the components of LUMO of **2a-1**., thereby raising HOMO significantly and LUMO slightly. Thereby, **2a-1** has a smaller $\Delta E_{\text{HOMO-LUMO}}$ than **2a**, which favors the shift of the electronic spectrum towards longer wavelengths. However, **2a-2** exhibits the exact opposite situation. This is due to diethylamino groups with alkyl cyclization, which limits the degree of free rotation of alkane in the substituent and also increases the number of C atoms in the p-electron delocalization of HOMO, thus favoring the shift of the electronic spectrum towards longer wavelengths. In addition, the $\Delta E_{\mathrm{HOMO-LUMO}}$ for three molecules (**2a-3** (2.50 eV), **2a-4** (2.65 eV) and **2a-5** (2.59 eV)) can also be minimized by further π-conjugation to some extent with respect to **2a-1**, **2a-2** and **2a**. Specifically, **2a-3** has the smallest $\Delta E_{\mathrm{HOMO-LUMO}}$, which is advantageous for achieving emission wavelengths within the biological transparent window range.

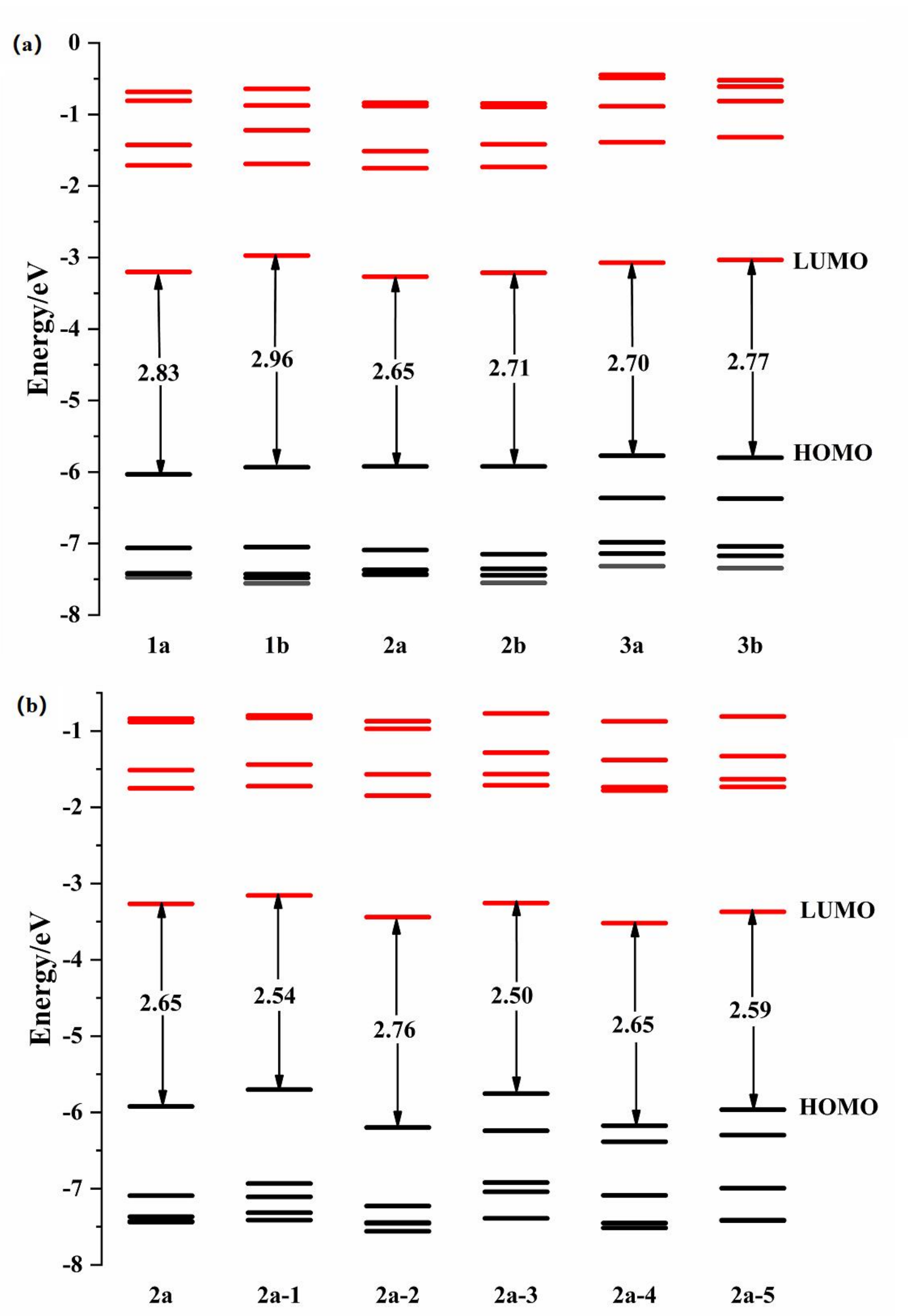

**Fig. 2** Calculated FMO energies of the studied complexes using DFT//B3LYP/6-31G(d, p). (a)the experimental molecules. (b)the designed molecules.

## 3.4. OPA and Emission Spectral Properties

Following the optimized $S_0$ geometries, the OPA and fluorescence emission properties are calculated, and the specific results along with the corresponding experimental values are summarized in **Table 1**. The simulated absorption and emission spectra of these molecules are drawn in **Fig. 3**. It is apparent that the maximum absorption and fluorescence emission peaks of all the studied molecules are mainly composed of HOMO→LUMO configuration. On the one hand, as shown in **Fig. 3(a)**, **a**-series molecules exhibit a significant red-shift in the maximum wavelength of OPA and fluorescence emission when compared to **b**-series molecules with the same substituted groups, and the corresponding oscillator strengths are slightly smaller than those of **b**-series molecules. it is worth noting that all the oscillator strengths of the two series of molecules are larger than 1, regardless of light absorbing or emitting process. Obviously, the difference in HOMO and LUMO between **a**-series and **b**-series molecules is responsible for this phenomenon. The addition of five-membered ring, which shortens the $O_1$-$C_2$ antibonding length in LUMO in **b**-series molecules, leading to its higher LUMO with respect to that of **a**-series. This consequently results in a comparatively wider HOMO-LUMO gap in **b**-series molecules, which accounts for the red-shift in electronic spectra of **a**-series molecules in comparison to **b**-series molecules. As can be seen in **Table 1**, the introduction of five-membered ring dramatically raises the LUMO level in **b**-series molecules, increasing the HOMO→LUMO transition energy and transition dipole moment. As a result, **b**-series molecules have a stronger oscillator strength than **a**-series molecules in terms of the formula (1) in **Supporting Information**. In addition, as compared to **b**-series molecules, the Stokes shifts of **a**-series molecules show an increasing trend. Fluorescent probes with larger Stokes shifts are more suitable for biological applications due to minimized self-quenching and fluorescence detection error arising from excitation backscattering effects.[40,41] As demonstrated in section **3.2**, this phenomenon is presumably because the six-membered rings make the **a**-series molecules geometrically more flexible and pliable than the **b**-series molecules when they are excited by light.

On the other hand, as shown in **Fig. 3(b)**, except for **2a-2**, the maximum OPA and fluorescence emission spectra wavelengths of all the designed molecules are red-shifted compared to **2a**. For **2a-1** the strong electron-donor capability of the $R_1$ substituent decreases the HOMO-LUMO gap, leading to a spectroscopic red-shift. For **2a-3~2a-5** obtained by further π-conjugation, the OPA and emission spectra are also red-shifted relative to **2a-1~2a-2** and **2a**, respectively. It is worth highlighting that **2a-3** has the longest emission wavelength (691.42 nm) and quite larger Stokes shifts (110.88 nm), which are more advantageous for bioimaging detection.

**Table 1** Calculated one-photon absorption and fluorescence emission properties of the studied molecules using B3LYP/6-311+G(d) method, including wavelength (λ), Stokes shift, vertical excitation energy (*E*), transition dipole moment ( *μ* ), oscillator intensity (*f*), transition characteristics, and corresponding experimental results.

| MOL. | Electronic transition | λ/nm | Stokes shift/nm | *E*/eV | *μ*/a.u. | *f* | Transition character | |
|---|---|---|---|---|---|---|---|---|
| 1a | $S_0 \rightarrow S_1$ | 519.95/542$^{expt}$ | 70.29/59$^{expt}$ | 2.38 | 4.62 | 1.2445 | H→L | 99.02% |
| | $S_1 \rightarrow S_0$ | 590.24/601$^{expt}$ | | 2.10 | 4.50 | 1.0430 | H→L | 99.66% |
| 1b | $S_0 \rightarrow S_1$ | 503.80/533$^{expt}$ | 66.17/57$^{expt}$ | 2.46 | 4.68 | 1.3224 | H→L | 99.10% |
| | $S_1 \rightarrow S_0$ | 569.97/590$^{expt}$ | | 2.18 | 4.60 | 1.1269 | H→L | 99.66% |
| 2a | $S_0 \rightarrow S_1$ | 551.59/574$^{expt}$ | 92.59/59$^{expt}$ | 2.25 | 5.05 | 1.4032 | H→L | 99.10% |
| | $S_1 \rightarrow S_0$ | 644.18/633$^{expt}$ | | 1.92 | 4.88 | 1.1251 | H→L | 99.76% |
| 2b | $S_0 \rightarrow S_1$ | 535.52/566$^{expt}$ | 84.42/54$^{expt}$ | 2.32 | 5.09 | 1.4698 | H→L | 99.24% |
| | $S_1 \rightarrow S_0$ | 619.94/620$^{expt}$ | | 2.00 | 4.96 | 1.2074 | H→L | 99.77% |
| **3a** | $S_0 \rightarrow S_1$ | 545.60/584$^{expt}$ | 69.01/55$^{expt}$ | 2.27 | 4.96 | 1.3689 | H→L | 98.61% |
| | $S_1 \rightarrow S_0$ | 614.61/639$^{expt}$ | | 2.02 | 4.68 | 1.0839 | H→L | 99.03% |
| **3b** | $S_0 \rightarrow S_1$ | 526.87/572$^{expt}$ | 66.13/48$^{expt}$ | 2.35 | 5.01 | 1.4499 | H→L | 98.57% |
| | $S_1 \rightarrow S_0$ | 593.00/620$^{expt}$ | | 2.09 | 4.86 | 1.2102 | H→L | 98.99% |
| **2a-1** | $S_0 \rightarrow S_1$ | 567.71 | 115.32 | 2.18 | 5.20 | 1.4493 | H→L | 99.26% |
| | $S_1 \rightarrow S_0$ | 683.03 | | 1.82 | 5.09 | 1.1535 | H→L | 100.00% |
| **2a-2** | $S_0 \rightarrow S_1$ | 538.90 | 70.86 | 2.30 | 5.00 | 1.4069 | H→L | 99.07% |
| | $S_1 \rightarrow S_0$ | 609.76 | | 2.03 | 4.86 | 1.1768 | H→L | 99.53% |
| **2a-3** | $S_0 \rightarrow S_1$ | 580.54 | 110.88 | 2.14 | 5.74 | 1.7264 | H→L | 97.70% |
| | $S_1 \rightarrow S_0$ | 691.42 | | 1.79 | 5.81 | 1.4820 | H→L | 99.35% |
| **2a-4** | $S_0 \rightarrow S_1$ | 570.69 | 96.67 | 2.17 | 4.86 | 1.2549 | H→L | 87.20% |
| | $S_1 \rightarrow S_0$ | 667.36 | | 1.86 | 1.38 | 1.0864 | H→L | 96.90% |
| **2a-5** | $S_0 \rightarrow S_1$ | 569.79 | 86.49 | 2.18 | 5.48 | 1.5986 | H→L | 94.21% |
| | $S_1 \rightarrow S_0$ | 656.28 | | 1.89 | 5.63 | 1.4664 | H→L | 97.21% |

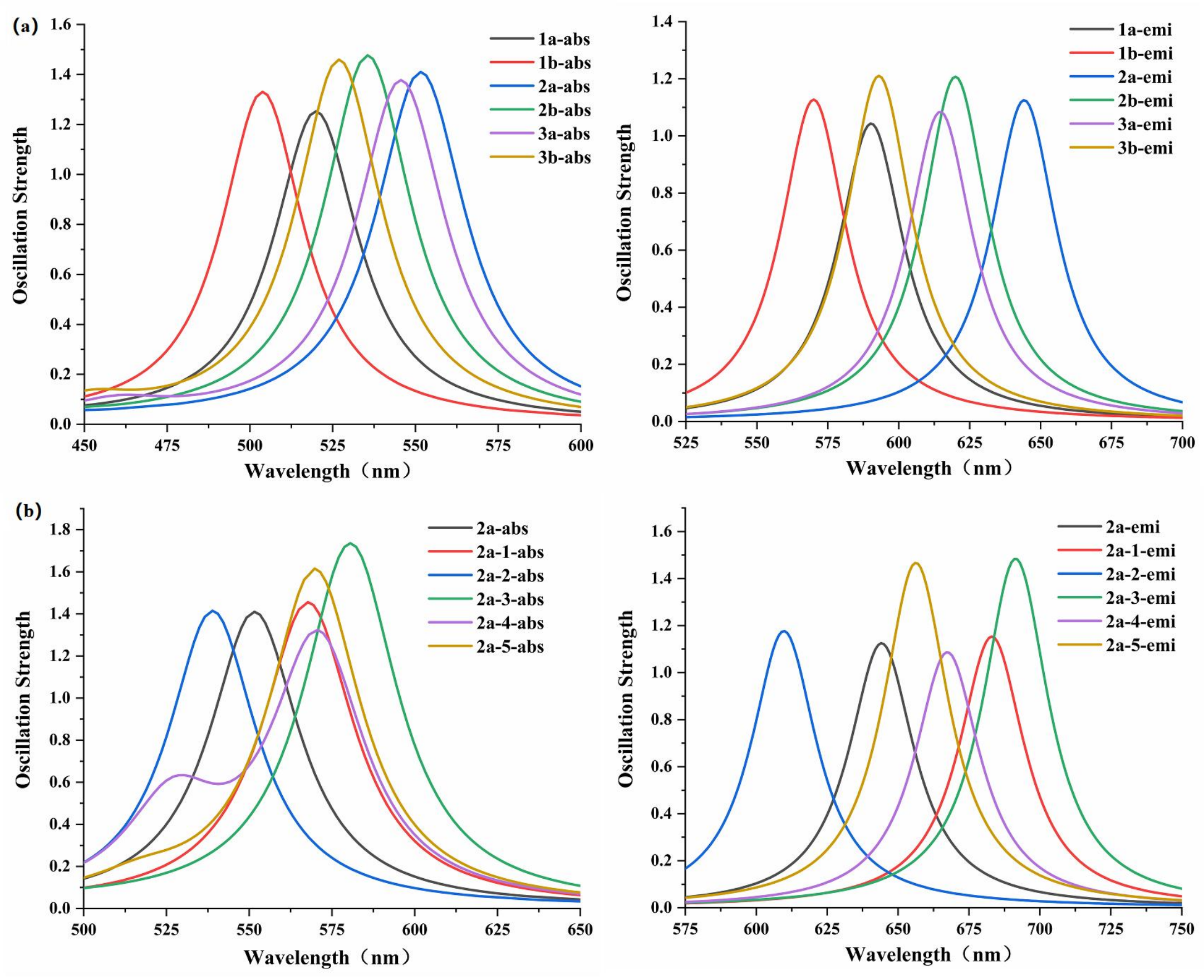

**Fig. 3** The simulated one-photon absorption and fluorescence emission spectra of the studied molecules, respectively. (a) the experimental molecules. (b). the designed molecules.

## 3.5. TPA Spectra Properties

As a two-photon biofluorescent probe and the corresponding product after recognizing the NTR reaction, it is necessary to have a large two-photon absorption cross-section, which implies a high photobleaching limits during the fluorescence detection process for NTR and thus ensures the lifetime of the probe. To evaluate the two-photon response of the studied molecules, the detailed physical parameters concerning the TPA spectra properties are listed in **Table 2**. The calculated maximum TPA cross section of **2a** is 185.73 GM, which is substantially consistent with the experimentally measured result (TPA cross section=89 GM/Φ=161.8 GM).[24] All of the studied compounds exhibit significant TPA peaks (110~785 GM) from 650 nm to 1100 nm, making them appropriate for bioimaging applications. The TPA wavelengths of **a**-series molecules are noticeably red-shifted (ca. 29~50 nm) in comparison to those of **b**-series molecules with the same substituted groups, yet the maximum TPA cross sections slightly decline (ca. 50~140 GM). As concerned designed molecules **2a-1**~**2a-5**, compared with **2a**, the TPA cross section values of the designed molecules **2a-3**, **2a-4** and **2a-5** are greatly improved by ca. 4~5 times. In particular, **2a-3**, whose TPA cross section reaches up 957.36 GM, makes it a promising candidate for high-performance two-photon fluorescent probes. However, how does the extending π conjugation or introducing subsitutents enhance TPA cross section values for these studied molecules? Herein, a generalized three states model (3SM) is adopted to analyze the molecular two-photon response, which can be described as follows:[42–45]

$$\delta_{TP}^{3SM} = \delta_{TP}^{ii} + \delta_{TP}^{ff} + 2\delta_{TP}^{if} \quad (3.1)$$

$$\delta_{TP}^{ii} = 8\left(\frac{\mu^{0i}\mu^{if}}{\Delta E_i}\right)^2 X^{ii} \quad (3.2)$$

$$\delta_{TP}^{ff} = 8\left(\frac{\mu^{0f}\mu^{ff}}{\Delta E_f}\right)^2 X^{ff} \quad (3.3)$$

$$\delta_{TP}^{if} = \sum_{i,f}\delta^{if} = 8\frac{\mu^{0i}\mu^{0f}\mu^{if}\mu^{ff}}{\Delta E_i \Delta E_f} X^{if} \quad (3.4)$$

where,

$$X^{ii} = (1 + 2\cos^2\theta_{0i}^{if}) \quad (3.5)$$

$$X^{ff} = (1 + 2\cos^2\theta_{0f}^{ff}) \quad (3.6)$$

$$X^{if} = \cos\theta_{0i}^{if}\cos\theta_{0f}^{ff} + \cos\theta_{0i}^{0f}\cos\theta_{if}^{ff} + \cos\theta_{0i}^{ff}\cos\theta_{0f}^{if} \quad (3.7)$$

in which, the ground state, the intermediate excited state, and the final state are noted by the indexes 0, *i*, and *f*, respectively. The transition dipole moments between the two states are represented by the $\mu$ terms. The difference in the excitation energies can be calculated using the following formula; $\Delta E_i = \omega_i - \frac{\omega_f}{2}$ , where $\omega_i$ means the excitation energy from the ground state to the *i* excited states. The angle terms ( $\theta_{0i}^{if}$ ) are obtained between the $\mu^{0i}$ and $\mu^{if}$ vectors. The $\delta_{TP}^{ii}$ terms always have positive values, However, owing to the dipole moments are relatively different in direction, the $\delta_{TP}^{if}$ terms exhibit two entirely different answers for positive and negative values, corresponding to the constructive and destructive interference effects, respectively. In other words, an increased or decreased TPA cross section could result from the relative distinct orientation between the various optical channels.

**Table 2** The TPA properties of the studied complexes calculated at B3LYP/6-311+G(d) level, including TPA spectra ($\lambda_{TPA}$), TPA cross-section ($\delta_{TPA}$), and the corresponding transition characters.

| MOL. | $\lambda_{TPA}$/nm | $\delta_{TPA}$/GM | Transition character | | |
|---|---|---|---|---|---|
| **1a** | 953.73 | 14.56 | $S_0 \rightarrow S_1$ | H→L | 99.02% |
| | 744.65 | 110.48 | $S_0 \rightarrow S_2$ | H-1→L | 92.48% |
| **1b** | 901.71 | 11.30 | $S_0 \rightarrow S_1$ | H→L | 99.10% |
| | 694.59 | 164.33 | $S_0 \rightarrow S_3$ | H-2→L | 93.36% |
| **2a** | 1008.01 | 27.66 | $S_0 \rightarrow S_1$ | H→L | 99.10% |
| | 746.90 | 185.73 | $S_0 \rightarrow S_2$ | H-1→L | 81.63% |
| **2b** | 980.12 | 22.60 | $S_0 \rightarrow S_1$ | H→L | 99.24% |
| | 714.61 | 323.74 | $S_0 \rightarrow S_2$ | H-1→L | 83.07% |
| **3a** | 999.88 | 36.45 | $S_0 \rightarrow S_1$ | H→L | 98.61% |
| | 731.47 | 181.85 | $S_0 \rightarrow S_3$ | H-2→L | 77.50% |
| | | | | H-3→L | 15.43% |
| **3b** | 968.63 | 24.82 | $S_0 \rightarrow S_1$ | H→L | 98.57% |
| | 704.46 | 236.01 | $S_0 \rightarrow S_3$ | H-2→L | 70.84% |
| | | | | H-3→L | 17.28% |
| **2a-1** | 1033.21 | 21.60 | $S_0 \rightarrow S_1$ | H→L | 99.26% |
| | 722.94 | 241.33 | $S_0 \rightarrow S_4$ | H-2→L | 69.56% |
| | | | | H-3→L | 13.22% |
| **2a-2** | 991.88 | 62.73 | $S_0 \rightarrow S_1$ | H→L | 99.07% |
| | 758.32 | 150.41 | $S_0 \rightarrow S_2$ | H-1→L | 77.75% |
| | | | | H-2→L | 17.21% |
| **2a-3** | 1059.70 | 11.47 | $S_0 \rightarrow S_1$ | H→L | 97.70% |
| | 751.42 | 957.36 | $S_0 \rightarrow S_4$ | H-2→L | 46.49% |
| | | | | H-3→L | 45.15% |
| **2a-4** | 1082.84 | 17.59 | $S_0 \rightarrow S_1$ | H→L | 87.20% |
| | | | | H-1→L | 12.26% |
| | 772.49 | 815.27 | $S_0 \rightarrow S_3$ | H-2→L | 90.34% |
| **2a-5** | 1050.72 | 11.14 | $S_0 \rightarrow S_1$ | H→L | 94.21% |
| | 765.34 | 792.13 | $S_0 \rightarrow S_3$ | H-2→L | 90.24% |
| **LDO-NTR** | 1271.64 | 0.13 | $S_0 \rightarrow S_1$ | H→L | 99.96% |
| | 692.65 | 166.19 | $S_0 \rightarrow S_5$ | H→L+2 | 98.50% |
| **2a-3-NTR** | 1467.28 | 0.27 | $S_0 \rightarrow S_1$ | H→L | 99.98% |
| | 882.46 | 210.56 | $S_0 \rightarrow S_4$ | H→L+2 | 95.99% |

As shown in **Fig. S12**, the calculated $\delta^{3SM}$ are largely in agreement with the results predicted by the response theory, indicating that the 3SM results can be applied for quantitative analysis of the relationship between molecular structures and TPA spectra properties. To begin with, the possible TPA transition channels are analyzed. **Table 2** illustrates that **1a**, **2a**, **2b**, and **2a-2** have only one kind of transition channel ($S_0 \rightarrow S_1 \rightarrow S_2$), since their maximum two-photon response occur at the $S_2$ states. Whereas, two possible transition channels ($S_0 \rightarrow S_1 \rightarrow S_3$ and $S_0 \rightarrow S_2 \rightarrow S_3$) may be involved in the TPA excitation processes for **1b**, **3a**, **3b**, **2a-4**, and **2a-5**, whose final TPA states are the $S_3$. As far as **2a-1** and **2a-3** are concerned, their higher TPA final states ($S_4$) lead to more complicated transition pathways. Subsequently, the relevant transition dipole moments of the compounds with multiple possible TPA transition channels are evaluated, and the results are given in **Table S8**. For **1b**, **3a**, and **3b**, we find that $\mu^{02}$ and $\mu^{23}$ are far smaller than $\mu^{01}$ and $\mu^{13}$, and thus the TPA cross sections, derived from a positive correlation with the product of $\mu^{02}$ and $\mu^{23}$ in the process of $S_0 \rightarrow S_2 \rightarrow S_3$, are also almost negligible. Consequently, it can be deduced that their mostly probable TPA transition channels are $S_0 \rightarrow S_1 \rightarrow S_3$. However, for all the designed molecules, the other possible transition channels also make a substantial contribution to the total TPA cross sections. Based on the above consideration, several important physical parameters of the designed molecules corresponding to all possible TPA transition channel are calculated. Subsequently, the excitation energies and transition dipole moments are provided in **Table 3**.

The detailed results illustrate that there is a relatively minor difference in the excitation energies between **a**-series and **b**-series molecules. To aid in visualizing the extent to which various dipole moment vectors and angle terms contribute to the TPA cross section, **Fig. S13** is plotted. The findings reveal that for both **a**-series and **b**-series of molecules, the transition dipole moments from the $S_0$ to intermediate state ($\mu^{0i}$) and the angle term of initial state ($X^{ii}$) are much larger than the other dipole moments and angle terms. The fact that $\delta^{ii}$ is positively correlated with the square of both $\mu^{0i}$ and $X^{ii}$ alongside explosive growth, as demonstrated in formula (**3.2**), also helps to explain why $\delta^{ii}$ is dominant in $\delta^{3SM}$. As shown in **Fig. S13(a)**, the **b**-series molecules display significant increases in their transition dipole moments ($\mu^{0i}$ and $\mu^{if}$), and state dipole moment ($\mu^{ff}$) when compared to the **a**-series molecules. This leads to a larger two-photon absorption, and thus rationalizing the **b**-series molecules own larger TPA cross sections. Indeed, $\mu^{if}$ plays a more crucial role in this process for the **b**-series molecules. In other words, the transition dipole moment from the intermediate state to the final state makes the compound bridged by five-membered ring, **b**-series molecules, more advantageous to increase TPA cross section. This cause is in line with the results of the FMOs analysis. Immediately following this, specific analysis of the three pairs of experimental molecules are performed. For both **1a** and **1b**, $X^{if}$ are greater than 0, while the values of **1b** exceed **1a**. Although $X^{ii}$ and $X^{ff}$ are marginally smaller for **1b**, the prominent dipole moments still leave **1b** with a larger TPA cross section. The cases of **3a** and **3b** are similar to the above discussion. As for **2a** and **2b**, they both have negative $X^{if}$ values, with the latter being smaller. However, the TPA cross section of **2b** prevails over that of **2a** due to the larger dipole moments and angle terms ($X^{ii}$ and $X^{ff}$).

Based on the **2a**, the **2a-1** and **2a-2** are obtained by internal cycloalkylation and external cyclization of the substituted amine groups, respectively. The **2a-3~2a-5** are designed by extending the π-conjugation, gaining better rigidity and considerably larger dipole moment vectors ($\mu^{0f}$, $\mu^{if}$ and $\mu^{ff}$) and decreased excitation energy (**Table 3**), which accounts for their higher TPA cross sections. Among a series of the designed molecules, **2a-3** has the most significant TPA cross section because of its largest transition dipole moment ($\mu^{0i}$) and smallest excitation energies ($\Delta E_i \Delta E_f$).

**Table 3** Selected important parameters about the 3SM process for all the studied molecules, including the excitation energies ($\Delta E$), the dipole moment vectors ($\mu$), the angle terms ($X$), and the selected $\delta$ components. (The $\mu$ and $\delta$ are units in a.u.)

| MOL. | Path | $\Delta E_i$ /a.u. | $\Delta E_f$ /a.u. | $\Delta E_i \Delta E_f$ /(a.u.)$^2$ | $\mu^{0i}$ | $\mu^{0f}$ | $\mu^{if}$ | $\mu^{ff}$ | $X^{ii}$ | $X^{ff}$ | $X^{if}$ | $\delta^{ii}$ ×10$^4$ | $\delta^{if}$ ×10$^4$ | $\delta^{3SM}$ ×10$^4$ |
|---|---|---|---|---|---|---|---|---|---|---|---|---|---|---|
| **1a** | $S_0 \rightarrow S_1 \rightarrow S_2$ | 0.03 | 0.06 | 0.0016 | 3.85 | 0.42 | 1.38 | 1.35 | 2.96 | 1.45 | 0.54 | 92.34 | 0.80 | 94.05 |
| **1b** | $S_0 \rightarrow S_1 \rightarrow S_3$ | 0.02 | 0.07 | 0.0016 | 3.92 | 0.27 | 1.59 | 2.21 | 2.92 | 1.02 | 0.58 | 157.26 | 1.07 | 159.47 |
| **2a** | $S_0 \rightarrow S_1 \rightarrow S_2$ | 0.02 | 0.06 | 0.0013 | 4.30 | 0.55 | 1.33 | 0.56 | 2.97 | 1.86 | -0.62 | 157.21 | -0.66 | 155.94 |
| **2b** | $S_0 \rightarrow S_1 \rightarrow S_2$ | 0.02 | 0.06 | 0.0014 | 4.34 | 0.72 | 1.68 | 1.72 | 2.98 | 2.85 | -0.92 | 262.14 | -4.79 | 253.44 |
| **3a** | $S_0 \rightarrow S_1 \rightarrow S_3$ | 0.02 | 0.06 | 0.0013 | 4.20 | 0.38 | 1.18 | 0.21 | 3.00 | 2.37 | -2.47 | 124.39 | -0.58 | 123.23 |
| **3b** | $S_0 \rightarrow S_1 \rightarrow S_3$ | 0.02 | 0.06 | 0.0014 | 4.25 | 0.86 | 1.34 | 2.91 | 2.69 | 1.06 | 0.20 | 136.69 | 1.59 | 141.18 |
| **2a-1** | $S_0 \rightarrow S_1 \rightarrow S_4$ | 0.02 | 0.06 | 0.0011 | 4.41 | 0.50 | 1.09 | 1.71 | 2.99 | 2.65 | 0.93 | 177.45 | 2.80 | 183.46 |
| | $S_0 \rightarrow S_2 \rightarrow S_4$ | 0.05 | 0.06 | 0.0032 | 0.48 | 0.50 | 1.11 | 1.71 | 2.23 | 2.65 | 0.42 | 0.19 | 0.05 | 0.69 |
| | $S_0 \rightarrow S_3 \rightarrow S_4$ | 0.06 | 0.06 | 0.0037 | 0.80 | 0.50 | 0.78 | 1.71 | 2.26 | 2.65 | -0.21 | 0.20 | -0.02 | 0.55 |
| **2a-2** | $S_0 \rightarrow S_1 \rightarrow S_2$ | 0.03 | 0.06 | 0.0015 | 4.39 | 0.41 | 1.18 | 0.39 | 2.99 | 1.10 | -0.01 | 102.06 | 0.00 | 102.06 |
| **2a-3** | $S_0 \rightarrow S_1 \rightarrow S_4$ | 0.02 | 0.06 | 0.0011 | 4.89 | 0.82 | 2.06 | 2.97 | 2.89 | 1.99 | -1.43 | 680.91 | -25.25 | 633.04 |
| | $S_0 \rightarrow S_2 \rightarrow S_4$ | 0.03 | 0.06 | 0.0018 | 0.57 | 0.82 | 1.00 | 2.97 | 2.99 | 1.99 | 0.25 | 0.84 | 0.15 | 3.76 |
| | $S_0 \rightarrow S_3 \rightarrow S_4$ | 0.05 | 0.06 | 0.0031 | 0.48 | 0.82 | 1.25 | 2.97 | 1.92 | 1.99 | -1.82 | 0.20 | -0.68 | 1.46 |
| **2a-4** | $S_0 \rightarrow S_1 \rightarrow S_3$ | 0.02 | 0.06 | 0.0013 | 2.78 | 1.25 | 0.70 | 2.39 | 2.17 | 1.00 | -0.31 | 13.48 | -1.12 | 13.40 |
| | $S_0 \rightarrow S_2 \rightarrow S_3$ | 0.03 | 0.06 | 0.0017 | 4.06 | 1.25 | 2.62 | 2.39 | 2.92 | 1.00 | -0.26 | 318.55 | -3.95 | 312.82 |
| **2a-5** | $S_0 \rightarrow S_1 \rightarrow S_3$ | 0.02 | 0.06 | 0.0013 | 3.74 | 0.99 | 1.29 | 2.07 | 2.68 | 1.36 | -0.80 | 108.75 | -5.08 | 99.95 |
| | $S_0 \rightarrow S_2 \rightarrow S_3$ | 0.03 | 0.06 | 0.0017 | 2.99 | 0.99 | 2.33 | 2.07 | 2.95 | 1.36 | 0.05 | 130.82 | 0.36 | 132.89 |

## 3.6. Theoretical validation of the fluorescence mechanism

The NTR-activated TP fluorescent probe **LDO-NTR** was experimentally observed to exhibit fluorescence enhancement with a 310-fold emission turn-on response upon the addition of NTR.[24] Furthermore, it was hypothesized that the negligible fluorescence of **LDO-NTR** originated from the combined effects of photoinduced electron transfer (PET) and internal charge transfer (ICT). However, to the best of our knowledge, the details about the fluorescence quenching mechanism still remain unclear. Obviously, it is of paramount importance of making clear the sensing mechanism used to recognize targets for more purposeful synthesis of probes in future. Thus, in the subsequent discussion, on the one hand, the fluorescence quenching mechanism of **LDO-NTR** is theoretically analyzed and confirmed from the following perspectives, including the frontier molecular orbitals (FMOs) theory, the natural population analysis (NPA), and the calculations of ICT and PET rates. On the other hand, for the **2a-3** molecule with the superior properties among all the designed molecules, the corresponding probe **2a-3-NTR** is designed and its fluorescence quenching mechanism is also predicted using the same method as described above

With the purpose of more accurately characterizing the properties of the probes and products, the experimental molecules (the probe **LDO-NTR** and product **2a**) are used as examples, and their one-photon absorption (OPA) and fluorescence emission properties are calculated at various functionals (TPSSH, B3LYP, PBE0, M06, BMK, M062X, CAM-B3LYP, wB97XD) and basis sets, and the detailed results are compiled in **Table S9-S10**. B3LYP/6-31G(d, p) is ultimately chosen as the optimal method for the subsequent calculations (Although emission peak is not quite accurate for probe **LDO-NTR** due to the self-interaction problem of B3LYP functional, the peak position of probe relative to its reaction product is not quite important for an Off-On probe). The OPA and emission properties of the probes and products are illustrated in **Table S11**, indicating that the transition from the $S_0$ to $S_4$ state for the probe **LDO-NTR** corresponds to the maximum absorption peak in the OPA process, which is composed of HOMO→LUMO+2. Meanwhile, the electron transitions from the $S_0$ to the first three excited states ($S_1$-$S_3$) of **LDO-NTR** with minimal oscillator strengths. From the frontier molecular orbitals depicted in **Fig. S14**, it can be noticed that the recognition group (4-nitrobenzyl alcohol) is not involved in the excitation process from the $S_0$ to the excited states, thus **LDO-NTR** is partially excited. In contrast, for the product **2a**, the transition from $S_0$ to $S_1$ with the largest oscillator strength is mainly involved by HOMO and LUMO, which are both delocalized throughout the molecule. The emission process almost experiences a similar situation, where the oscillator strength of $S_1$ state is zero, i.e. transition probability is zero, the radiative transition in the $S_1$→$S_0$ process is completely inhibited. The excited state $S_1$ then returns to the ground state in a nonradiative way. Besides, the designed probe **2a-3-NTR** molecule has the similar case to **LDO-NTR**.

Now the question is whether the nonradiative decays occur by PET process or ICT process or both? And why does it happen in this particular system? Initially, the Rhem-Weller formula, which reads $\Delta G_{ET} = E^0(D^+/D) - E^0(A/A^-) - \Delta E_{0,0} + w$, must be satisfied for PET to occur. In the following discussion, the detailed calculations for physical parameters such as $E^0(D^+/D)$, $E^0(A/A^-)$, and $\Delta E_{0,0}$, etc are given. The results showed that the probes have a rather strong driving force ($-\Delta G_{ET}$) in the PET process (**LDO-NTR**: -0.3806 eV, **2a-3-NTR**: -0.3139 eV), while the corresponding reaction products fail to meet this point. Sequentially, in order to address the question of why it happens in this system, the activated fluorophore components (anthocyanin) and recognition groups (4-nitrobenzyl alcohol) of the probes are called as Flu* and Receptor, respectively.Their frontier molecular orbital energy levels are calculated on the basis of the whole molecule,[46–48] with the results displayed in **Fig. 4**. For the probe **LDO-NTR**, the HOMO energy of the Receptor is -2.62 eV, which is higher than the Flu* (-5.05 eV). Therefore, upon excitation of **LDO-NTR**, i.e., after the Flu* is excited, an electron can be transferred from the HOMO of the Receptor to the HOMO of the Flu* by Rhem-Weller formula. And an electron lying in the LUMO of Flu* cannot instantaneously back to HOMO of Flu* itself, resulting in the quenching of the fluorescence, that is, photoinduced electron transfer (PET) takes place. In contrast, for the product **2a**, the hydroxyl group in the product **2a** has a small twist angle relative to the entire molecule, so the electron cloud distributions of the HOMO and LUMO cannot be completely separated and therefore are delocalized throughout the whole molecule, which inhibit the PET process, and a dramatic fluorescence phenomenon occur in **2a**. As can be seen in **Fig. 4**, the designed probe **2a-3-NTR** and product **2a-3** exhibit similar behavior to the experimental molecules, suggesting that **2a-3-NTR** is also involved in the PET process.

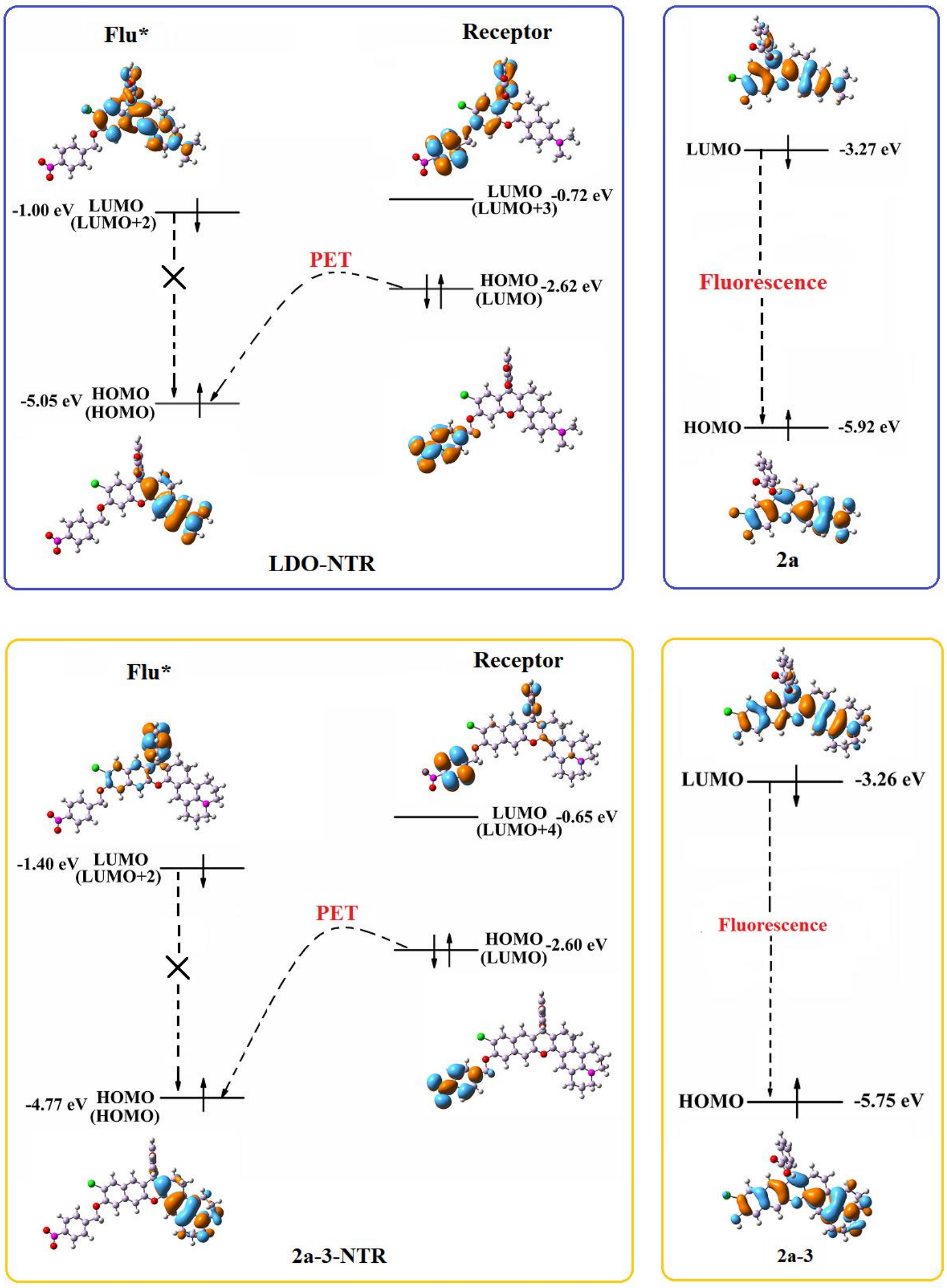

**Fig. 4** Theoretical validation of the PET mechanism for the studied molecules.

In addition, the possibility of the ICT mechanism was proposed in the recognition reaction of NTR by the fluorescent probe **LDO-NTR**, accompanied by the PET process, which will also be theoretically unravelled below. As shown in **Fig. S15**, each probe (**LDO-NTR** and **2a-3-NTR**) and product (**2a** and **2a-3**) is divided into three fragments, and their natural population analysis (NPA) are presented in **Table S12**. Evidently, the recognition group (4-nitrobenzyl alcohol) in the probes functions as an electron acceptor. After the reaction with NTR, the 4-nitrobenzyl alcohol group is hydrolyzed to generate a phenolic compound, with the hydroxyl group serving as an electron donor. Analyzing the differences in natural charge population (ΔQ) between the $S_1$ and $S_0$ states of each probe's fragment (ΔQ: fragment I<0, II>0, III>0), indicates that the negative charges increase in fragment I and decrease in II and III. While the products have the opposite effect (ΔQ: fragment I>0, II<0, III>0), meaning that the negative charges decrease in fragment I and III and increase in II. Thus, it may be inferred that intramolecular charge transfer (ICT) is involved in the excited state relaxation process of all the probes and products.

The above discussion simply demonstrates the existence of the PET and ICT mechanisms by the FMOs theory and NPA analysis. An oxidation-reduction reaction dyad is formed between the receptor and fluorophore in the probe because it has an extra receptor component than the corresponding product. The Rhem-Weller formula reveals that an electron in higher HOMO level of receptor would be transferred to HOMO of the excited fluorophore under the perturbation interaction of its hole-electron pair. Because of the sufficiently large negative free energy, this redox reaction can occur spontaneously. As a result, the PET process happens in the probe. The next questions are, what are the reaction rates if the probes undergo PET and ICT processes? Do the fluorophores still have enough fluorescence in their excited states to compete with the PET? From the following Marcus equation,[49–51] the rates are determined to further confirm the occurrence of the two mechanisms.

$$k = \sqrt{\frac{4\pi^3}{h^2 \lambda k_B T}} V^2 \exp\left[ -\frac{\left(\Delta G^0 + \lambda\right)^2}{4\lambda k_B T} \right] \tag{3.8}$$

where Planck's constant and Boltzmann constant are represented by $h$ and $k_B$, respectively. The absolute temperature ($T$) is set to 298 K. $\lambda$ and $\Delta G^0$ stand for reorganization energy and free energy change, respectively. The symbol $V$ denotes the electronic coupling matrix element.

The Rehm-Weller equation[52,53] can be utilized to calculate the reaction free energy ($\Delta G_{ET}$) in the PET process.

$$\Delta G_{ET} = E^0\left(D^+ / D\right) - E^0\left(A / A^-\right) - \Delta E_{0,0} + w \tag{3.9}$$

where the oxidation potential of the electron-donor, $E^0(D^+/D)$, is defined as the energy difference between its optimized ionic and neutral states. Likewise, $E^0(A/A^-)$ represents the energy difference between the optimized neutral and ionic states of the electron-acceptor fragment. As shown in **Fig. S15-S16**, the probes are separated into donor and acceptor segments based on the findings listed in **Table S12**. The ICT and PET process share the same precise divisions for the probes. The zero-zero transition energy is denoted by $\Delta E_{0,0}$. One way to estimate the $w$ term in the formula is to:

$$w = \frac{1}{4\pi\varepsilon_0} \cdot \frac{e^2}{\varepsilon R} \tag{3.10}$$

where the dielectric constants in the vacuum and water solvent are represented by $\varepsilon_0$ and $\varepsilon$, respectively. The electron charge is referred to $e$, and the distance between the donor and acceptor is symbolized by R.

The following formula can be used to determine the reorganization energy in the PET process:[54]

$$\lambda \approx \frac{e^2}{4\pi\varepsilon_0}\left(\frac{1}{2r_1} + \frac{1}{2r_2} - \frac{1}{R}\right)\left(\frac{1}{n^2} - \frac{1}{\varepsilon}\right) \tag{3.11}$$

in which $r_1$ and $r_2$ denote the radii of the electron donor and acceptor, respectively. The symbol n corresponds to the refractive index of the water solvent.

The expression of $\Delta G^0$ during the ICT process is as follows:[55]

$$\Delta G_{\mathrm{CT}} = E_{\mathrm{EA}}(\mathrm{A}) - E_{\mathrm{IP}}(\mathrm{D}) - \Delta E_{0\text{-}0} - E_{\mathrm{b}} \tag{3.12}$$

in this case, the ionization potential of the donor and electron affinity of the acceptor, respectively, are represented by $E_{\mathrm{IP}}(\mathrm{D})$ and $E_{\mathrm{EA}}(\mathrm{A})$. The energy of the lowest excited state of free donor is marked by $\Delta E_{0\text{-}0}$, while the exciton binding energy of the whole molecule is expressed as $E_{\mathrm{b}}$.

The following equation performs the reorganization energy in the ICT process:

$$\lambda = E\left(\mathrm{A}^-\right) - E(\mathrm{A}) + E(\mathrm{D}) - E\left(\mathrm{D}^+\right) \tag{3.13}$$

here, $E(\mathrm{A}^-)$ and $E(\mathrm{A})$ are the energies of the neutral acceptor at anionic and optimal ground state geometries, respectively. $E(\mathrm{D})$ and $E(\mathrm{D}^+)$ note the energies of the radical cation donor at neutral and optimal cation geometries, respectively.

Ultimately, the Generalized Mulliken-Hush method[56] can be used to estimate the electronic coupling term $V$ of these two rate equations.

$$V = \frac{\mu_{12}\Delta E_{12}}{\sqrt{\Delta\mu_{12}^2 + 4\mu_{12}^2}} \tag{3.14}$$

Where, $\mu_{12}$ is the electron transition dipole moment between the initial state and final state. The dipole moment difference and the energy difference between the two states, are symbolized by $\Delta\mu_{12}$ and $\Delta E_{12}$, respectively.

The aforementioned equations are used to calculate the PET and ICT rates based on the $S_0 \rightarrow S_1$ process in this paper, and the theoretically simulated related physical parameters are provided in **Table 4**. It can be seen that the ICT rate for **LDO-NTR** is $1.02 \times 10^{14}$ $s^{-1}$, which is noticeably faster than the PET rate ($7.36 \times 10^{9}$ $s^{-1}$). Stated otherwise, the probe **LDO-NTR** may effortlessly transition from a local excited state to an ICT state,that is, charge redistribution, after which it can either radiate a photon for fluorescence or undergo PET through electron transfer. As a result of the fluorescence process being forbidden (the oscillator strength of $S_1 \rightarrow S_0$ is zero, see **Table S11**), ICT happens upon light excitation and PET occurs immediately thereafter. Simultaneously, the situation remains almost exactly similar for the designed probe **2a-3-NTR**. Consequently, the negligible fluorescence of the probes results from the combined effects of PET and ICT. These findings unmistakably show the dynamic trend of the PET and ICT processes. Hence, it is possible to draw the following conclusions: (1) Since the fluorescence emission process is prohibited for **LDO-NTR**, PET occurs between the fluorophore and receptor, whether by calculating the kinetic rates or by estimating their states using the Rhem-Weller formula. After **LDO-NTR** reacts with NTR, the HOMO and LUMO electron clouds are delocalized throughout the product **2a**. Subsequently, the PET process is interrupted, resulting in a strong fluorescence. Furthermore, both the probe **LDO-NTR** and product **2a** involve the ICT process. (2) The designed probe **2a-3-NTR** might share the same fluorescence quenching mechanism as **LDO-NTR**.

**Table 4** Calculated results about the PET and ICT rates for the two type probes, including $V$ is the electronic coupling term, λ is the reorganization energy, and $k$ is the rate.

| MOL. | ICT | | | | | |
|---|---|---|---|---|---|---|
| | $V$/eV | $\Delta G_{CT}$/eV | $\lambda_{CT}$/eV | $\Delta E_{0-0}$/eV | $E_b$/eV | $k_{ICT}$/s$^{-1}$ |
| **LDO-NTR** | 0.8047 | 0.0392 | 0.4438 | 2.27 | 0.22 | $1.02\times10^{14}$ |
| **2a-3-NTR** | 0.0848 | -0.2298 | 0.4463 | 2.29 | 0.18 | $6.53\times10^{13}$ |
| **MOL.** | **PET** | | | | | |
| | R/A | $V$/eV | $\Delta G_{ET}$/eV | $\lambda_{ET}$/eV | $w$/eV | $k_{PET}$/s$^{-1}$ |
| **LDO-NTR** | 9.59 | 0.8047 | 0.3806 | 0.3663 | 0.0192 | $7.36\times10^{9}$ |
| **2a-3-NTR** | 11.45 | 0.0848 | 0.3139 | 0.4501 | 0.0160 | $6.56\times10^{8}$ |

## 3.7. Solvation Free Energy

Bearing in mind that two-photon fluorescent probes are usually used for biological tissues and thus, good solubility is essential. As listed in **Table 5**, the solvation free energy ($\Delta G_{solv}$) of all the studied molecules are calculated in water solution. The detailed results indicate that the $\Delta G_{solv}$ of the designed probe **2a-3-NTR** (-19.50 kcal/mol) and product **2a-3** (-55.04 kcal/mol) remain almost comparable to those of the experimental probe **LDO-NTR** (-18.46 kcal/mol) and product **2a** (-56.16 kcal/mol). This implies that the design strategy of replacing the dialkylamino group by alkyl cyclization and expanding the π-conjugation fails to worsen the water solubility for **LDO-NTR**-based probes.

**Table 5** Calculated solvation free energy $\Delta G_{solv}$ (kcal/mol) of the studied molecules.

| MOL. | $\Delta G_{solv}$（kcal/mol） |
|---|---|
| **LDO-NTR** | -18.46 |
| **2a** | -56.16 |
| **2a-3-NTR** | -19.50 |
| **2a-3** | -55.04 |

## 4. Conclusions

In this paper, a series of anthocyanidins with different backbones and substituents are chosen and designed as NTR probes, and their photophysical properties are thoroughly investigated. Subtle differences in photophysical nature resulting from non-conjugated modifications in their substituents and backbones are fully revealed. The research on a series of anthocyanidins with the same substituents but different fluorophore skeletons **1a-3a** and **1b-3b** clarifies that the effect of six/five-membered rings in the backbone on their photophysical properties. The results indicate that six/five-membered rings fused in the backbone have considerable differences in deep electronic occupied orbitals and geometrical spatial hindrance upon excitation. In comparison with the five-membered ring fused in the backbone, the six-membered ring in the backbone makes molecular geometry free relaxation. However, **a**-series molecules have a higher reorganization energy, generating more energy loss upon light excitation, which allows the reaction products can detect NTR by a larger Stokes shift. More importantly, the loss of fluorescence intensity is extremely low when the Stokes shift is increased. These features are highly useful for high-resolution NTR detection. On this basis, we have designed **2a-n** (**n**=**1**-**5**) series probe molecules, by using different substituent and backbone π congjugation modifications to improve their two-photon absorption and fluorescence performance. As a result, **2a-3** has a better figure of merit: its emission wavelength, Stokes shift and TPA cross section are as high as 691.42 nm, 110.88 nm and 957.36 GM, respectively, making its probe molecule **2a-3-NTR** a promising candidate for high-performance two-photon fluorescence probes. In addition, we also demonstrate how the experimental and designed anthocyanin probes quench fluorescence, and the involvement of ICT and PET mechanisms in the NTR detection are elucidated and validated by using static state and dynamic rate. The findings show that after being excited by light, both the probes and the products reach an excited state, and the ICT process occurs instantaneously. The probe molecule then undergoes a PET process, causing fluorescence quenching, while the product emit photons from the excited state and then return to the $S_0$ state due to the lack of PET structural feature. Given the larger Stokes shift (110.88 nm), two-photon absorption cross section (957.36 GM), and longer fluorescence wavelength (691.42 nm), the designed two-photon fluorescence probe molecule **2a-3-NTR** is considered the most promising candidate molecule for detecting NTR. Further experimental confirmation is expected.

## Author contributions

**Xiu-e Zhang**: Investigation, Data curation, Writing–original draft. **Xue Wei**: Methodology, Writing–review & editing. **Wei-Bo Cui**: Writing–review & editing. **Jin-Pu Bai**: Writing–review & editing. **Aynur Matyusup**: Writing–review & editing. **Jing-Fu Guo**: Software. **Hui Li**: Project administration, Funding acquisition. **Ai-Min Ren**: Supervision, Writing–review & editing.

## Conflicts of interest

There are no conflicts to declare.

## Acknowledgments

This work was supported by Natural Science Foundation of Jilin Province of China (No.20240101167JC), the Key Research and Development Project of Jilin Provincial

Department of Science and Technology (No.20240302015GX), the 2020-JCJQ project (GFJQ2126-007) and the Natural Science Foundation of China (nos. 21473071, 21173099, and 20973078).

## 5.References

Supporting Information for

# Rational Designing of Anthocyanidins-Directed Near-Infrared Two-Photon Fluorescence Probes

Xiu-e Zhang[a], Xue Wei[b], Wei-Bo Cui[b], Jin-Pu Bai[a], Aynur Matyusup[a], Jing-Fu Guo*[a], Hui Li[b] and Ai-Min Ren*[b]

[a]School of Physics, Northeast Normal University, Changchun 130024, P.R.China

[b]Laboratory of Theoretical and Computational Chemistry, Institute of Theoretical Chemistry, College of Chemistry, Jilin University, Liutiao Road #2, Changchun 130061, P.R.China

Corresponding Author: Jing-Fu Guo

E-mail: guojf217@nenu.edu.cn

ORCID: 0000-0002-1864-167X

Corresponding Author: Ai-Min Ren

E-mail: renam@jlu.edu.cn

ORCID: 0000-0002-9192-1483

# Table of Contents

## Theoretical Approaches

## Electronic structures and photophysical properties

13. **Table S10** Calculated one-photon absorption and fluorescent emission spectra properties of the probe **2a** and the product **LDO-NTR** by using different basis sets and B3LYP functional.
14. **Table S11** Calculated one-photon absorption and fluorescence emission properties of the studied probes and products molecules by using B3LYP/6-31G(d, p).
15. **Table S12** The change of natural charge population (ΔQ/e) of each fragment of the studied molecules during the transition, including g stands for the ground state, and e represents the excited state.
16. **Fig. S1** The chemical structures and atomic numbering of the molecules (**1a**, **1b** and **1c**).
17. **Fig. S2** The superimposed structures of the $S_0$ and $S_1$ states and the corresponding RMSD values for the studied molecules. (the red and the blue represent the $S_0$ and $S_1$ states, respectively.)
18. **Fig. S3** The main frontier molecular orbitals of the studied experimental molecules.
19. **Fig. S4** The main frontier molecular orbitals of the molecules **1a**, **1b** and **1c**.
20. **Fig. S5** The FMO energies of the molecules (**1a**, **1b** and **1c**) by DFT//B3LYP/6-31G(d, p).
21. **Fig. S6** The simulated one-photon absorption and fluorescence emission spectra of the studied molecules (**1a**, **1b** and **1c**).
22. **Fig. S7** The chemical structures of the molecules **2a**~**5a**.
23. **Fig. S8** The main frontier molecular orbitals of the molecules **2a**~**5a**.
24. **Fig. S9** The FMO energies of the studied complexes (**2a**~**5a**) by DFT//B3LYP/6-31G(d, p).
25. **Fig. S10** The simulated one-photon absorption and fluorescence emission spectra of the studied molecules (**2a**~**5a**).
26. **Fig. S11** The main frontier molecular orbitals of the designed molecules.
27. **Fig. S12** Comparison between the $\delta^{3SM}$ calculated by using the three-state model and the $\delta^{TPA}$ predicted by the response theory of the studied molecules. Besides, the $\delta^{ii}$ is the first item in the three-state model formula. (a)the experimental molecules. (b)the designed molecules.

28. **Fig. S13** Plots of (a) the dipole moment vectors μ and (b) the angle terms *X* existed in the 3SM of the studied experimental molecules.
29. **Fig. S14** The main frontier molecular orbitals of the studied probes and products.
30. **Fig. S15** Detailed division of the probes and products. (I in green, II in black, III in red)
31. **Fig. S16** The detailed division of electron donor and electron acceptor regions for the probe **LDO-NTR** and **2a-3-NTR**, including the green is the electron acceptor and the red is the electron donor.

# Theoretical Methods

## 1. One-Photon Absorption (OPA)

The OPA transition intensity is calculated by the following formula:[1]

$$\delta_{\mathrm{OP}} = \frac{2\omega_{\mathrm{f}}}{3} \sum_{\alpha} \left| \langle 0 | \mu_{\alpha} | f \rangle \right|^2 \tag{1}$$

Here, $\omega_f$ is the excitation energy from the ground state $S_0$ to the singlet excited state $S_f$, $\mu_\alpha$ is the electron transition dipole moment and the sum of the *x*, *y* and *z* directions of the molecule.

## 2. Two-Photon Absorption (TPA)

The TPA cross section ($\delta_{\mathrm{TP}}$) determines the TPA intensity and is associated with the TPA transition probability, which can be obtained from the two-photon transition matrix element ($S_{\alpha\beta}$). And the expression of $S_{\alpha\beta}$ is as follows:[2]

$$S_{\alpha\beta} = \sum_{i} \left[ \frac{\langle 0 | \mu_{\alpha} | i \rangle \langle i | \mu_{\beta} | f \rangle}{\omega_i - \frac{\omega_f}{2}} + \frac{\langle 0 | \mu_{\beta} | i \rangle \langle i | \mu_{\alpha} | f \rangle}{\omega_i - \frac{\omega_f}{2}} \right] \tag{2}$$

Where $\mu_\alpha$ and $\mu_\beta$ are dipole moment operators, respectively. $\omega_i$ represents the excitation energy from $S_0$ to $S_i$, and $\omega_f / 2$ represents half of the excitation energy from $S_0$ to $S_f$.

Then, with the help of $S_{\alpha\beta}$, the TPA transition probability can be defined by the following equation:[3]

$$\delta_{\mathrm{a.u.}} = \frac{1}{30} \sum_{\alpha,\beta} \left( 2 S_{\alpha\alpha} S^{*}_{\beta\beta} + 4 S_{\alpha\beta} S^{*}_{\beta\alpha} \right) \tag{3}$$

Finally, the $\delta_{\mathrm{TP}}$ value can be carried out as follows:[4]

$$\delta_{\mathrm{TP}} = \frac{4\pi^2 \mathrm{a}_0^5 \alpha \omega^2}{15\mathrm{c}\Gamma} \delta_{\mathrm{au}} \tag{4}$$

Here, $\alpha$ and $a_0$ represent the fine structure parameter and Bohr radius, respectively. $\omega$ and $c$ stand for the photon $a_0$ energy in atomic units and the speed of light, respectively. $\Gamma$ refers to the broadening factor of the final state energy level, which is assumed to be 0.1 eV.[5]

## 3. Solvation Free Energy

The solvation free energy is the difference between the solute free energy in solution and gas phase, which can be calculated by the following formula: [6,7]

$$\Delta G_{solv} = G_{solv} - G_{gas} \tag{5}$$

Here, $G_{\mathrm{solv}}$ and $G_{\mathrm{gas}}$ are the energy of the molecule in the solvent and the gas phase, respectively.

**Table S1** Calculated one-photon absorption and fluorescent emission spectra properties of **1a** and **1b** by using different functionals and 6-31G(d, p) basis set.

| **Functionals** | **1a** | | | | **1b** | | | |
|---|---|---|---|---|---|---|---|---|
| | $\lambda_{\mathbf{max,abs}}$ | $f^{o}$ | $\lambda_{\mathbf{max,ems}}$ | $f^{e}$ | $\lambda_{\mathbf{max,abs}}$ | $f^{o}$ | $\lambda_{\mathbf{max,ems}}$ | $f^{e}$ |
| **TPSSH** | 523.02 | 1.1646 | 608.35 | 0.9113 | 494.88 | 1.2496 | 566.50 | 0.9895 |
| **B3LYP** | 508.31 | 1.2163 | 575.73 | 1.0207 | 492.46 | 1.2924 | 555.22 | 1.1038 |
| **PBE0** | 495.05 | 1.2585 | 555.73 | 1.0889 | 480.13 | 1.3407 | 538.42 | 1.1678 |
| **M06** | 494.20 | 1.2382 | 551.17 | 1.0915 | 481.71 | 1.3074 | 536.71 | 1.1538 |
| **BMK** | 473.71 | 1.3473 | 520.19 | 1.2304 | 448.98 | 1.4346 | 492.21 | 1.3116 |
| **M062X** | 471.02 | 1.3648 | 516.53 | 1.2739 | 456.02 | 1.4559 | 502.96 | 1.3505 |
| **CAM-B3LYP** | 464.85 | 1.3609 | 514.66 | 1.2735 | 449.38 | 1.4585 | 495.50 | 1.3690 |
| **wB97XD** | 460.64 | 1.3576 | 505.55 | 1.2927 | 446.86 | 1.4526 | 488.70 | 1.3805 |
| **EXP.** | 542 | | 601 | | 533 | | 590 | |

**Table S2** Calculated one-photon absorption and fluorescent emission spectra properties of **1a** and **1b** by using different basis sets and B3LYP functional.

| **Basis sets** | **1a** | | | | **1b** | | | |
|---|---|---|---|---|---|---|---|---|
| | $\lambda_{\mathbf{max,abs}}$ | $f^{o}$ | $\lambda_{\mathbf{max,ems}}$ | $f^{e}$ | $\lambda_{\mathbf{max,abs}}$ | $f^{o}$ | $\lambda_{\mathbf{max,ems}}$ | $f^{e}$ |
| **6-31G(d,p)** | 508.31 | 1.2163 | 575.73 | 1.0207 | 492.46 | 1.2924 | 555.22 | 1.1038 |
| **6-31+G(d)** | 517.07 | 1.2455 | 586.54 | 1.0429 | 501.04 | 1.3235 | 566.31 | 1.1268 |
| **6-311G(d,p)** | 513.70 | 1.2260 | 582.38 | 1.0289 | 497.67 | 1.3024 | 562.08 | 1.1117 |
| **6-311+G(d)** | 519.95 | 1.2445 | 590.24 | 1.0430 | 503.80 | 1.3224 | 569.97 | 1.1269 |
| **6-311++G(d,p)** | 519.45 | 1.2461 | 588.91 | 1.0447 | 490.67 | 1.3456 | 568.74 | 1.1289 |
| **EXP.** | 542 | | 601 | | 533 | | 590 | |

**Table S3** The one-photon absorption and fluorescent emission spectra properties of all the experimental molecules were calculated by using 6-311+G(d) basis set and B3LYP/TPSSH functionals.

| **MOL.** | **Method** | **Electronic transition** | **λ/nm** | **Stokes shift/nm** | ***E*/eV** | ***f*** | **Configuration** | |
|---|---|---|---|---|---|---|---|---|
| **1a** | **B3LYP** | $S_0 \rightarrow S_1$ | 519.95/542$^{expt}$ | 70.29/59$^{expt}$ | 2.38 | 1.2445 | H→L | 99.02% |
| | | $S_1 \rightarrow S_0$ | 590.24/601$^{expt}$ | | 2.10 | 1.0430 | H→L | 99.66% |
| | **TPSSH** | $S_0 \rightarrow S_1$ | 533.63/542$^{expt}$ | 88.10/59$^{expt}$ | 2.32 | 1.1930 | H→L | 99.30% |
| | | $S_1 \rightarrow S_0$ | 621.73/601$^{expt}$ | | 1.99 | 0.9330 | H→L | 100% |
| **1b** | **B3LYP** | $S_0 \rightarrow S_1$ | 503.80/533$^{expt}$ | 66.17/57$^{expt}$ | 2.46 | 1.3224 | H→L | 99.10% |
| | | $S_1 \rightarrow S_0$ | 569.97/590$^{expt}$ | | 2.18 | 1.1269 | H→L | 99.66% |
| | **TPSSH** | $S_0 \rightarrow S_1$ | 504.21/533$^{expt}$ | 74.69/57$^{expt}$ | 2.46 | 1.2815 | H→L | 99.30% |
| | | $S_1 \rightarrow S_0$ | 578.90/590$^{expt}$ | | 2.14 | 1.0094 | H→L | 99.83% |
| **2a** | **B3LYP** | $S_0 \rightarrow S_1$ | 551.59/574$^{expt}$ | 92.59/59$^{expt}$ | 2.25 | 1.4032 | H→L | 99.10% |
| | | $S_1 \rightarrow S_0$ | 644.18/633$^{expt}$ | | 1.92 | 1.1251 | H→L | 99.76% |
| | **TPSSH** | $S_0 \rightarrow S_1$ | 572.09/574$^{expt}$ | 116.96/59$^{expt}$ | 2.17 | 1.3304 | H→L | 99.53% |
| | | $S_1 \rightarrow S_0$ | 689.05 /633$^{expt}$ | | 1.80 | 0.9932 | H→L | 100% |
| **2b** | **B3LYP** | $S_0 \rightarrow S_1$ | 535.52/566$^{expt}$ | 84.42/54$^{expt}$ | 2.32 | 1.4698 | H→L | 99.24% |
| | | $S_1 \rightarrow S_0$ | 619.94/620$^{expt}$ | | 2.00 | 1.2074 | H→L | 99.77% |
| | **TPSSH** | $S_0 \rightarrow S_1$ | 555.60/566$^{expt}$ | 82.17/54$^{expt}$ | 2.23 | 1.3862 | H→L | 99.50% |
| | | $S_1 \rightarrow S_0$ | 637.77/620$^{expt}$ | | 1.94 | 1.0678 | H→L | 100% |
| **3a** | **B3LYP** | $S_0 \rightarrow S_1$ | 545.60/584$^{expt}$ | 69.01/55$^{expt}$ | 2.27 | 1.3689 | H→L | 98.61% |
| | | $S_1 \rightarrow S_0$ | 614.61/639$^{expt}$ | | 2.02 | 1.0839 | H→L | 99.03% |
| | **TPSSH** | $S_0 \rightarrow S_1$ | 558.95/584$^{expt}$ | 102.14/55$^{expt}$ | 2.22 | 1.3240 | H→L | 97.91% |
| | | $S_1 \rightarrow S_0$ | 661.09/639$^{expt}$ | | 1.88 | 0.9351 | H→L | 99.41% |
| **3b** | **B3LYP** | $S_0 \rightarrow S_1$ | 526.87/572$^{expt}$ | 66.13/48$^{expt}$ | 2.35 | 1.4499 | H→L | 98.57% |
| | | $S_1 \rightarrow S_0$ | 593.00/620$^{expt}$ | | 2.09 | 1.2102 | H→L | 98.99% |
| | **TPSSH** | $S_0 \rightarrow S_1$ | 547.74/572$^{expt}$ | 82.10/48$^{expt}$ | 2.26 | 1.3421 | H→L | 97.10% |
| | | $S_1 \rightarrow S_0$ | 629.84/620$^{expt}$ | | 1.97 | 1.0878 | H→L | 99.20% |

**Table S4** Calculated two-photon absorption properties of **2a** by using B3LYP/6-311+G(d) and CAM-B3LYP/6-311+G(d) method.

| MOL. | Method | $\lambda_{TPA}$/nm | $\delta_{TPA}$/GM | Transition character | | |
|---|---|---|---|---|---|---|
| 2a | B3LYP/6-311+G(d) | 1008.01 | 27.66 | $S_0 \rightarrow S_1$ | H→L | 99.10% |
| | | 746.90 | 185.73/161.8[expt] | $S_0 \rightarrow S_2$ | H-1→L | 81.63% |
| 2a | CAM-B3LYP/6-311+G(d) | 928.73 | 34.90 | $S_0 \rightarrow S_1$ | H→L | 94.61% |
| | | 635.82 | 581.00/161.8[expt] | $S_0 \rightarrow S_2$ | H-1→L | 75.56% |

**Table S5** Some bond lengths (Å) and dihedral angles (deg) at the optimized $S_0$ and $S_1$ states for all the studied molecules.

| MOL. | $O_1$-$C_2$ | $C_2$-$C_3$ | $C_3$-$C_{10}$ | $C_{10}$-$C_n$[a] | $C_2$-$C_n$[b] | DHA1[c] | DHA2[d] | DHA3[e] | DHA4[f] | DHA5[g] |
|---|---|---|---|---|---|---|---|---|---|---|
| 1a-$S_0$ | 1.3422 | 1.4194 | 1.4291 | 2.9125 | 1.4221 | -148.02 | 71.27 | -12.92 | 1.81 | -10.77 |
| 1a-$S_1$ | 1.3643 | 1.4612 | 1.4240 | 2.9238 | 1.3778 | -151.26 | 66.94 | -14.70 | 5.80 | -12.80 |
| 1b-$S_0$ | 1.3297 | 1.4123 | 1.4255 | 2.3640 | 1.4130 | 177.62 | 78.32 | 0.38 | 3.55 | 0.81 |
| 1b-$S_1$ | 1.3514 | 1.4456 | 1.4196 | 2.3724 | 1.3774 | 175.07 | 57.41 | 0.32 | 8.88 | 2.39 |
| 1c-$S_0$ | 1.3465 | 1.4267 | 1.4221 | 3.0120 | 1.4092 | - | 61.55 | 0.67 | 4.43 | 1.36 |
| 1c-$S_1$ | 1.3650 | 1.4714 | 1.4146 | 3.0379 | 1.3627 | - | 50.34 | 2.71 | 10.16 | 3.74 |
| 2a-$S_0$ | 1.3442 | 1.4152 | 1.4295 | 2.9134 | 1.4259 | -147.73 | 71.76 | -12.49 | 2.33 | -10.28 |
| 2a-$S_1$ | 1.3616 | 1.4654 | 1.4164 | 2.9216 | 1.3766 | -151.22 | 69.09 | -15.16 | 5.51 | -13.68 |
| 2b-$S_0$ | 1.3300 | 1.4049 | 1.4260 | 2.3675 | 1.4220 | 178.08 | 66.24 | -0.23 | 2.26 | 0.60 |
| 2b-$S_1$ | 1.3470 | 1.4503 | 1.4125 | 2.3722 | 1.3761 | 175.11 | 56.60 | 0.31 | 7.54 | 2.25 |
| 3a-$S_0$ | 1.3451 | 1.4214 | 1.4263 | 2.9138 | 1.4172 | -149.18 | 112.11 | -12.80 | -1.20 | -11.73 |
| 3a-$S_1$ | 1.3668 | 1.4661 | 1.4175 | 2.9239 | 1.3736 | -146.18 | 121.44 | -17.20 | -3.12 | -17.10 |
| 3b-$S_0$ | 1.3305 | 1.4093 | 1.4243 | 2.3677 | 1.4168 | 177.35 | 66.06 | 0.06 | 0.51 | 1.01 |
| 3b-$S_1$ | 1.3512 | 1.4494 | 1.4139 | 2.3746 | 1.3760 | 174.75 | 55.47 | 0.48 | 1.31 | 2.64 |
| 2a-1-$S_0$ | 1.3488 | 1.4104 | 1.4332 | 2.9305 | 1.4274 | -147.04 | 72.62 | -13.37 | 2.22 | -10.14 |
| 2a-1-$S_1$ | 1.3600 | 1.4664 | 1.4149 | 2.9367 | 1.3749 | -149.53 | 67.09 | -15.99 | 6.20 | -14.17 |
| 2a-2-$S_0$ | 1.3396 | 1.4243 | 1.4234 | 2.9132 | 1.4189 | 146.83 | 70.43 | 12.95 | 2.54 | 11.29 |
| 2a-2-$S_1$ | 1.3638 | 1.4622 | 1.4190 | 2.9243 | 1.3778 | 147.40 | 62.82 | 15.67 | 6.46 | 15.22 |
| 2a-3-$S_0$ | 1.3438 | 1.4089 | 1.4342 | 2.9311 | 1.4326 | -147.41 | 74.23 | -13.06 | 1.74 | -9.78 |
| 2a-3-$S_1$ | 1.3543 | 1.4621 | 1.4167 | 2.9355 | 1.3809 | -149.53 | 70.66 | -16.05 | 4.53 | -14.08 |
| 2a-4-$S_0$ | 1.3345 | 1.4211 | 1.4257 | 2.9125 | 1.4267 | -146.95 | 69.86 | -12.62 | 2.05 | -10.55 |
| **2a-4-S1** | 1.3780 | 1.4291 | 1.4237 | 2.9222 | 1.4069 | -152.06 | 80.82 | -13.71 | 2.82 | -10.93 |
| **2a-5-S0** | 1.3390 | 1.4136 | 1.4305 | 2.9138 | 1.4318 | -148.49 | 73.00 | -12.36 | 1.81 | -10.15 |
| **2a-5-S1** | 1.3574 | 1.4563 | 1.4192 | 2.9186 | 1.3859 | -151.34 | 73.87 | -14.93 | 3.62 | -13.14 |

[a]: $C_n$ refers to $C_{13}$ in **a**-series molecules, $C_{12}$ in **b**-series molecules, respectively.

[b]: $C_n$ refers to $C_{13}$ in **a**-series molecules, $C_{12}$ in **b**-series molecules and $C_{11}$ in **c**-series molecules, respectively.

[c]: DHA1 is the dihedral angle between the atoms 6-13-12-11 (**a**-series molecules) or 6-12-11-10 (**b**-series molecules).

[d]: DHA2 is the dihedral angle between the atoms 5-6-7-8.

[e]: DHA3 is the dihedral angle between the atoms 13-2-3-10 (**a**-series molecules) or 12-2-3-10 (**b**-series molecules) or 11-2-3-10 (**c**-series molecules).

[f]: DHA4 is the dihedral angle between the atoms 6-7-8-9.

[g]: DHA5 is the dihedral angle between the atoms 1-2-3-4.

**Table S6** Calculated reorganization energy (λ) of the studied molecules, in which $E_{S1}$ and $E_{S0}$ represent the energies of the excited and ground states, respectively.

| MOL. | $E_{S1}$/Hartree | $E_{S0}$/Hartree | λ/cm$^{-1}$ |
|---|---|---|---|
| **1a** | -1280.9625 | -1280.9683 | 1262.04 |
| **1b** | -1241.6442 | -1241.6496 | 1187.46 |
| **1c** | -1203.5270 | -1203.5326 | 1240.38 |
| **2a** | -1819.1649 | -1819.1711 | 1370.09 |
| **2b** | -1779.8440 | -1779.8501 | 1338.33 |
| **3a** | -1745.1281 | -1745.1332 | 1101.87 |
| **3b** | -1705.8060 | -1705.8109 | 1089.32 |

**Table S7** The one-photon absorption and fluorescence emission properties of the studied molecules (**2a**~**5a**) calculated by B3LYP/6-311+G(d) method, including wavelength (λ), Stokes shift, vertical excitation energy (*E*), transition dipole moment (*μ*), oscillator intensity (*f*), transition character.

| **MOL.** | **Electronic transition** | **λ/nm** | **Stokes shift/nm** | ***E*/eV** | ***μ*/a.u.** | ***f*** | **Transition character** | |
|---|---|---|---|---|---|---|---|---|
| **1a** | $S_0 \rightarrow S_1$ | 519.95/542[expt] | 70.29/59[expt] | 2.38 | 4.62 | 1.2445 | H→L | 99.02% |
| | $S_1 \rightarrow S_0$ | 590.24/601[expt] | | 2.10 | 4.50 | 1.0430 | H→L | 99.66% |
| **1b** | $S_0 \rightarrow S_1$ | 503.80/533[expt] | 66.17/57[expt] | 2.46 | 4.68 | 1.3224 | H→L | 99.10% |
| | $S_1 \rightarrow S_0$ | 569.97/590[expt] | | 2.18 | 4.60 | 1.1269 | H→L | 99.66% |
| **1c** | $S_0 \rightarrow S_1$ | 520.19 | 71.78 | 2.38 | 4.62 | 1.2448 | H→L | 99.07% |
| | $S_1 \rightarrow S_0$ | 591.97 | | 2.09 | 4.38 | 0.9831 | H→L | 99.61% |
| **2a** | $S_0 \rightarrow S_1$ | 551.59/574[expt] | 92.59/59[expt] | 2.25 | 5.05 | 1.4032 | H→L | 99.10% |
| | $S_1 \rightarrow S_0$ | 644.18/633[expt] | | 1.92 | 4.88 | 1.1251 | H→L | 99.76% |
| **3a** | $S_0 \rightarrow S_1$ | 545.60/584[expt] | 69.01/55[expt] | 2.27 | 4.96 | 1.3689 | H→L | 98.61% |
| | $S_1 \rightarrow S_0$ | 614.61/639[expt] | | 2.02 | 4.68 | 1.0839 | H→L | 99.03% |
| **4a** | $S_0 \rightarrow S_1$ | 523.63 | 67.02 | 2.37 | 4.76 | 1.3121 | H→L | 99.17% |
| | $S_1 \rightarrow S_0$ | 590.65 | | 2.10 | 4.66 | 1.1144 | H→L | 99.71% |
| **5a** | $S_0 \rightarrow S_1$ | 547.54 | 80.75 | 2.26 | 4.90 | 1.3347 | H→L | 98.98% |
| | $S_1 \rightarrow S_0$ | 628.29 | | 1.97 | 4.72 | 1.0792 | H→L | 99.64% |

**Table S8** The related transition dipole moments (units in a.u.) of the studied molecules with multiple possible TPA transition channels.

| MOL. | $\mu^{01}$ | $\mu^{13}$ | $\mu^{02}$ | $\mu^{23}$ | | |
|---|---|---|---|---|---|---|
| **1b** | 3.92 | 1.59 | 0.44 | 0.63 | | |
| **3a** | 4.20 | 1.18 | 0.93 | 0.38 | | |
| **3b** | 4.25 | 1.34 | 1.06 | 0.90 | | |
| **2a-4** | 2.78 | 0.70 | 4.06 | 2.62 | | |
| **2a-5** | 3.74 | 1.29 | 2.99 | 2.33 | | |
| **MOL.** | $\mu^{01}$ | $\mu^{14}$ | $\mu^{02}$ | $\mu^{24}$ | $\mu^{03}$ | $\mu^{34}$ |
| **2a-1** | 4.41 | 1.09 | 0.48 | 1.11 | 0.80 | 0.78 |
| **2a-3** | 4.89 | 2.06 | 0.57 | 1.00 | 0.48 | 1.25 |

**Table S9** Calculated one-photon absorption and fluorescent emission spectra properties of the probe **2a** and the product **LDO-NTR** by using different functionals and 6-31G(d, p) basis set.

| Functionals | 2a | | | | LDO-NTR | | | |
|---|---|---|---|---|---|---|---|---|
| | $\lambda_{max,abs}$ | $f^{O}$ | $\lambda_{max,ems}$ | $f^{e}$ | $\lambda_{max,abs}$ | $f^{O}$ | $\lambda_{max,ems}$ | $f^{e}$ |
| **TPSSH** | 513.00 | 1.0298 | 674.45 | 0.9817 | 720.20 | 0.0000 | 1414.91 | 0.0000 |
| **B3LYP** | 540.17 | 1.3932 | 629.14 | 1.1129 | 558.41 | 0.0000 | 938.54 | 0.0000 |
| **PBE0** | 526.96 | 1.4422 | 605.16 | 1.1949 | 471.00 | 0.0001 | 731.68 | 0.0000 |
| **M06** | 524.85 | 1.4271 | 594.13 | 1.2125 | 462.24 | 0.0001 | 698.14 | 0.0000 |
| **BMK** | 499.46 | 1.5549 | 555.55 | 1.3811 | 362.65 | 0.0004 | 508.06 | 0.0000 |
| **M062X** | 494.89 | 1.5763 | 544.92 | 1.4467 | 309.39 | 0.0000 | 558.32 | 0.0000 |
| **CAM-B3LYP** | 486.53 | 1.5768 | 541.48 | 1.4551 | 307.04 | 0.0000 | 549.43 | 0.0000 |
| **wB97XD** | 482.25 | 1.5774 | 531.61 | 1.4837 | 306.61 | 0.0000 | 547.15 | 0.0000 |
| **EXP.** | 574 | | 633 | | 561 | | 624 | |

**Table S10** Calculated one-photon absorption and fluorescent emission spectra properties of the probe **2a** and the product **LDO-NTR** by using different basis sets and B3LYP functional.

| Basis sets | 2a | | | | LDO-NTR | | | |
|---|---|---|---|---|---|---|---|---|
| | $\lambda_{max,abs}$ | $f^{O}$ | $\lambda_{max,ems}$ | $f^{e}$ | $\lambda_{max,abs}$ | $f^{O}$ | $\lambda_{max,ems}$ | $f^{e}$ |
| **6-31G(d,p)** | 540.17 | 1.3932 | 629.14 | 1.1129 | 558.41 | 0.0000 | 938.54 | 0.0000 |
| **6-31+G(d)** | 549.09 | 1.4031 | 641.10 | 1.1233 | 626.85 | 0.0000 | 1125.54 | 0.0000 |
| **6-311G(d,p)** | 546.28 | 1.4053 | 637.04 | 1.1247 | 565.64 | 0.0000 | 952.46 | 0.0000 |
| **6-311+G(d)** | 551.59 | 1.4032 | 644.18 | 1.1251 | 628.03 | 0.0000 | 1126.72 | 0.0000 |
| **6-311++G(d,p)** | 551.69 | 1.4034 | 644.17 | 1.1251 | 628.87 | 0.0000 | 1128.24 | 0.0000 |
| **EXP.** | 574 | | 633 | | 561 | | 624 | |

**Table S11** Calculated one-photon absorption and fluorescence emission properties of the studied probes and products molecules by using B3LYP/6-31G(d, p).

| **Molecules** | **Electronic transition** | ***λ*/nm** | ***E*/eV** | ***f*** | **Configuration** | |
|---|---|---|---|---|---|---|
| Absorption | | | | | | |
| **LDO-NTR** | $S_0 \rightarrow S_1$ | 558.41/561$^{expt}$ | 2.22 | 0.0000 | H→L | 99.96% |
| | $S_0 \rightarrow S_2$ | 392.56 | 3.16 | 0.0078 | H→L+1 | 98.22% |
| | $S_0 \rightarrow S_3$ | 383.42 | 3.23 | 0.0042 | H-1→L | 97.37% |
| | $S_0 \rightarrow S_4$ | 346.49 | 3.58 | 0.7225 | H→L+2 | 98.36% |
| **2a** | $S_0 \rightarrow S_1$ | 540.17/574$^{expt}$ | 2.30 | 1.3932 | H→L | 99.21% |
| **2a-3-NTR** | $S_0 \rightarrow S_1$ | 624.64 | 1.98 | 0.0000 | H→L | 99.98% |
| | $S_0 \rightarrow S_2$ | 443.48 | 2.80 | 0.2085 | H→L+1 | 98.48% |
| **2a-3** | $S_0 \rightarrow S_1$ | 571.12 | 2.17 | 1.7092 | H→L | 97.09% |
| Emission | | | | | | |
| **LDO-NTR** | $S_1 \rightarrow S_0$ | 938.54/624$^{expt}$ | 1.32 | 0.0000 | H→L | 99.97% |
| **2a** | $S_1 \rightarrow S_0$ | 629.14/633$^{expt}$ | 1.97 | 1.1129 | H→L | 99.92% |
| **2a-3-NTR** | $S_1 \rightarrow S_0$ | 1129.69 | 1.10 | 0.0000 | H→L | 99.98% |
| **2a-3** | $S_1 \rightarrow S_0$ | 676.49 | 1.83 | 1.4677 | H→L | 99.35% |

**Table S12.** The change of natural charge population (ΔQ/e) of each fragment of the studied molecules during the transition, including g stands for the ground state, and e represents the excited state.

| **MOL.** | | **I** | **II** | **III** |
|---|---|---|---|---|
| **LDO-NTR** | $\mathbf{Q_g}$ | -0.1906 | 0.1731 | 0.0175 |
| | $\mathbf{Q_e}$ | -1.1458 | 0.8246 | 0.3212 |
| | **ΔQ** | -0.9552 | 0.6515 | 0.3038 |
| **2a** | $\mathbf{Q_g}$ | -0.1376 | 0.9907 | 0.1469 |
| | $\mathbf{Q_e}$ | -0.1308 | 0.8779 | 0.2528 |
| | **ΔQ** | 0.0068 | -0.1128 | 0.1060 |
| **2a-3-NTR** | $\mathbf{Q_g}$ | -0.1968 | 0.1152 | 0.0817 |
| | $\mathbf{Q_e}$ | -1.1555 | 0.6931 | 0.4624 |
| | **ΔQ** | -0.9587 | 0.5779 | 0.3807 |
| **2a-3** | $\mathbf{Q_g}$ | -0.1576 | 0.3515 | 0.8061 |
| | $\mathbf{Q_e}$ | -0.1443 | 0.2684 | 0.8759 |
| | **ΔQ** | 0.0133 | -0.0831 | 0.0698 |

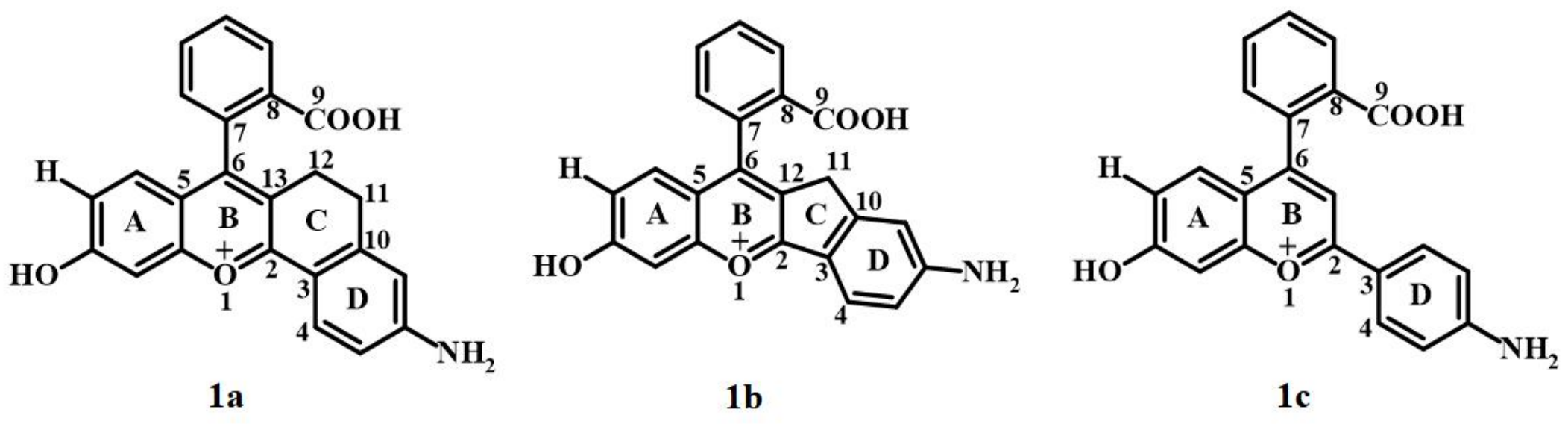

**Fig. S1** The chemical structures and atomic numbering of the molecules (**1a**, **1b** and **1c**).

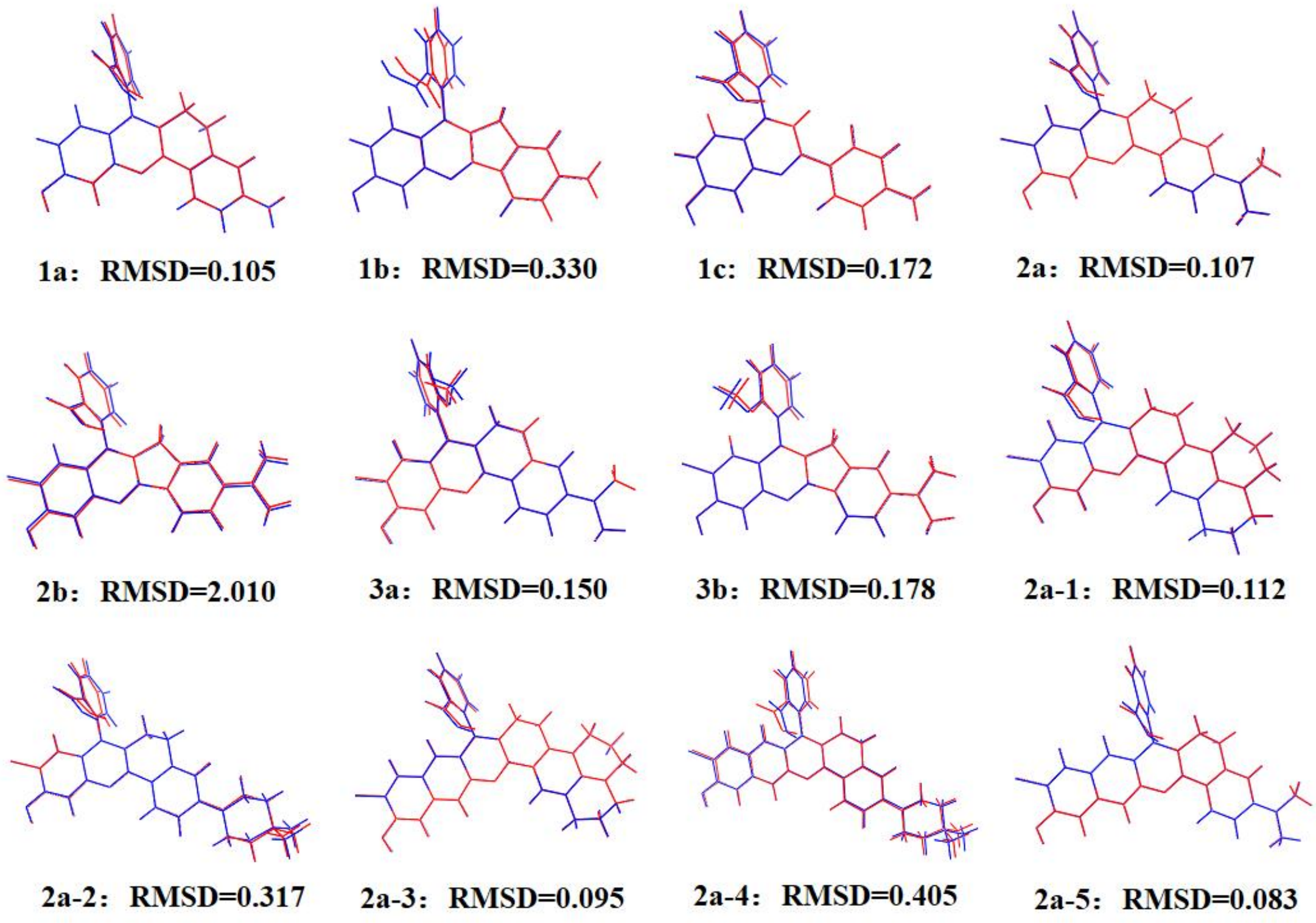

**Fig. S2** The superimposed structures of the $S_0$ and $S_1$ states and the corresponding RMSD values for the studied molecules. (the red and the blue represent the $S_0$ and $S_1$ states, respectively.)

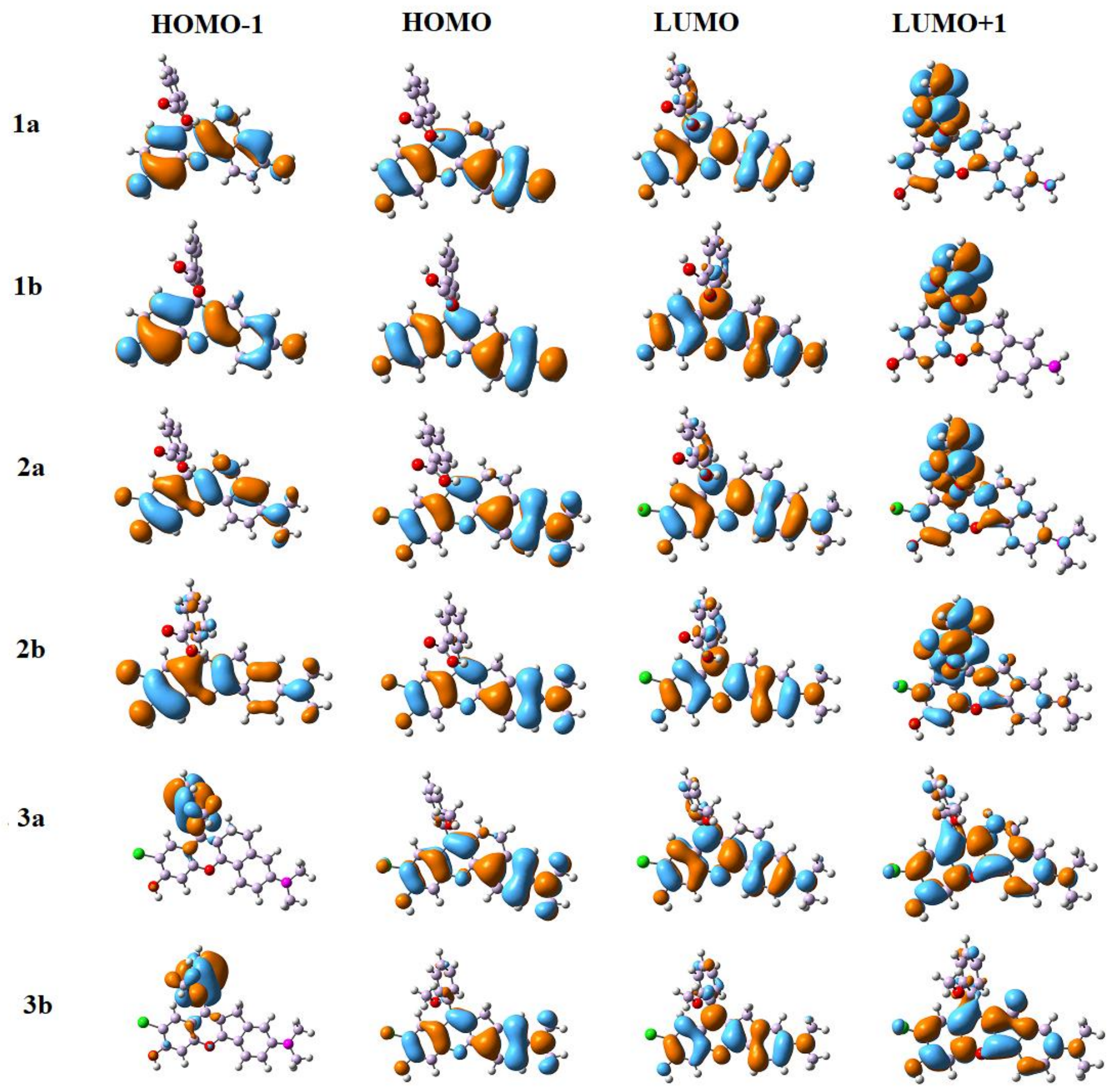

**Fig. S3** The main frontier molecular orbitals of the studied experimental molecules.

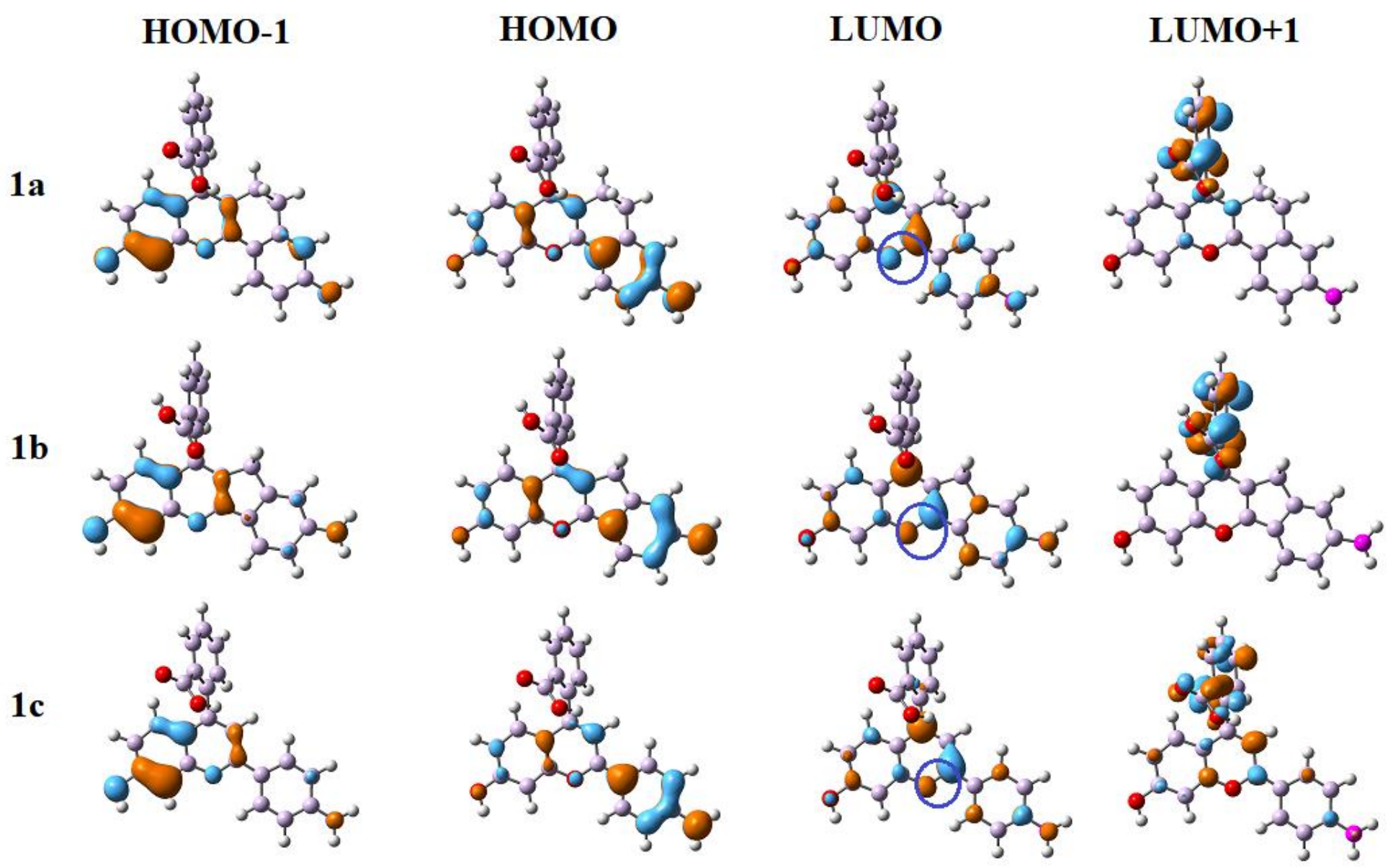

**Fig. S4** The main frontier molecular orbitals of the molecules **1a**, **1b** and **1c**.

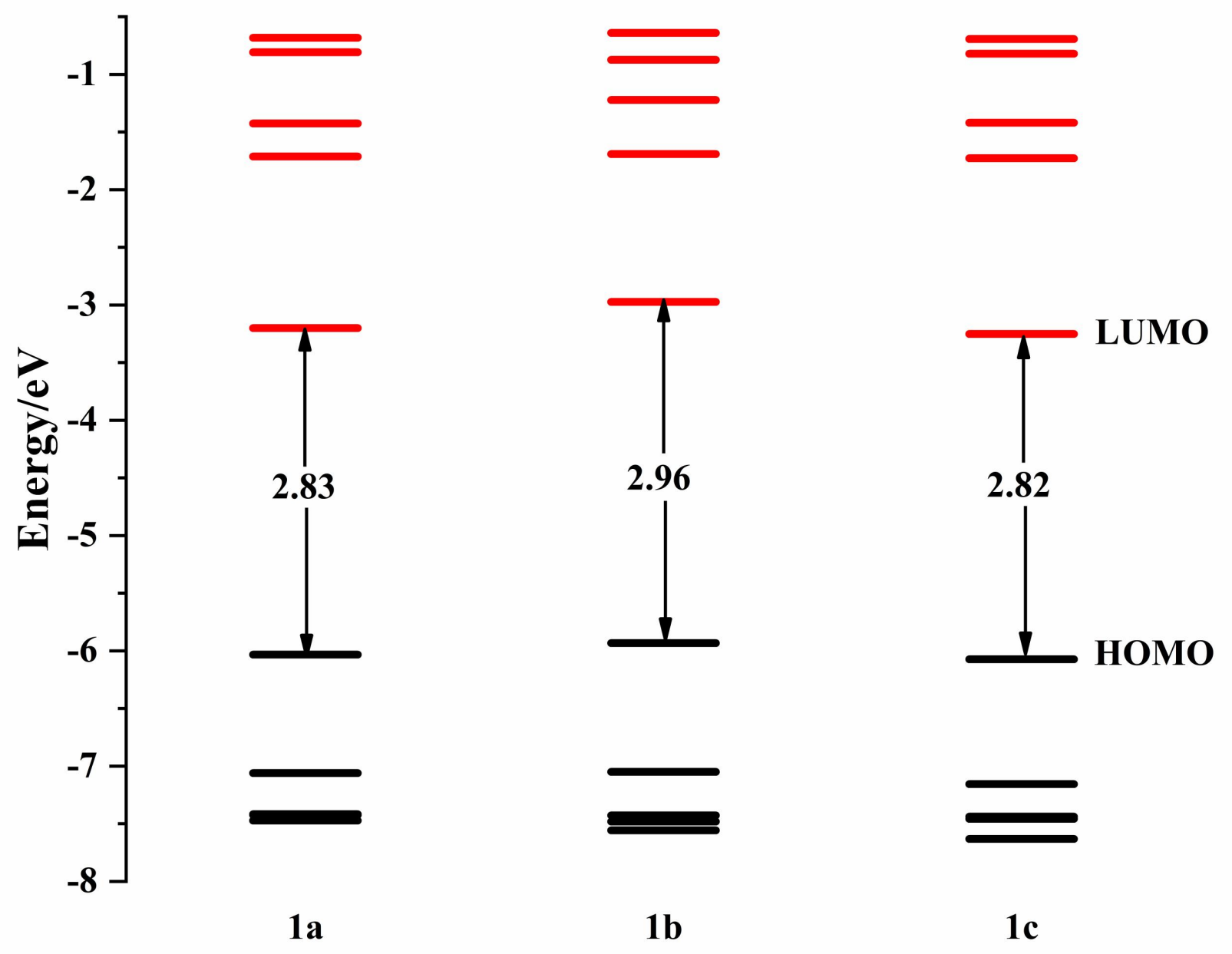

**Fig. S5** The FMO energies of the molecules (**1a**, **1b** and **1c**) by DFT//B3LYP/6-31G(d, p).

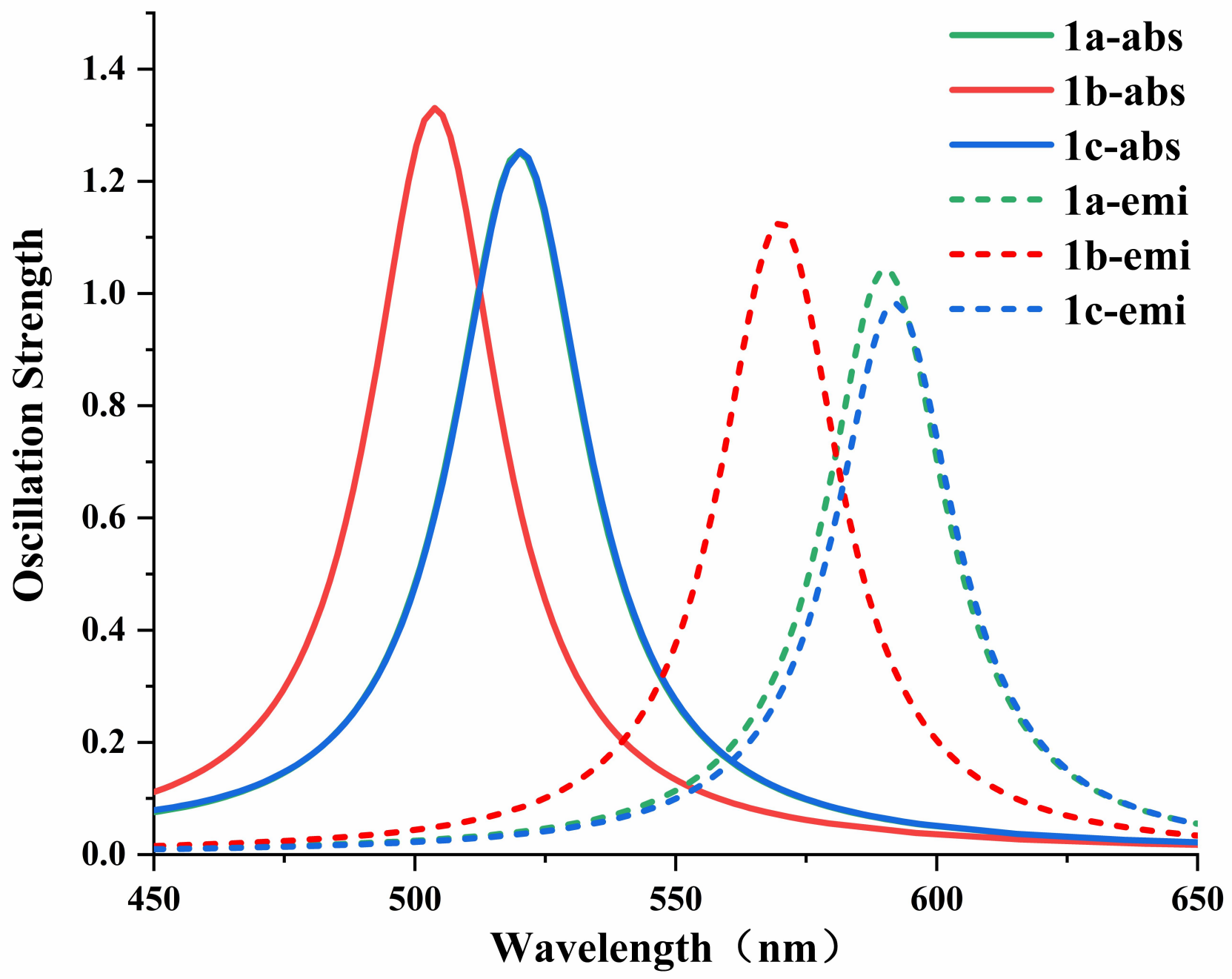

**Fig. S6** The simulated one-photon absorption and fluorescence emission spectra of the studied molecules (**1a**, **1b** and **1c**).

In order to find suitable molecular design strategies for the purpose of improving the photophysical properties of **2a** molecules (such as emission wavelength, Stokes shift and TPA cross sections), the **2a**~**5a** molecules with different substituents are investigated and their chemical structures are shown in **Fig. S7**. The calculated frontier molecular orbitals and the corresponding energy levels are shown in **Fig. S8** and **Fig. S9**. The results indicate that for **2a** and **4a** molecules with different $R_1$ substituents, the former has a significantly smaller HOMO-LUMO gap owing to the fact that $-N(CH_3)_2$ has a stronger electron-donating capacity than $-NH_2$ group, which in turn leads to a higher HOMO energy levels. It's observed that **2a** and **3a** with different $R_3$ substituents have the similar situation as the above discussion. And for **2a** and **5a** molecules with different $R_2$ substituents, since -Cl has a greater electron-withdrawing capacity than -H, which make **2a** has a lower LUMO energy level, ultimately resulting in a smaller $\Delta E_{HOMO\text{–}LUMO}$. Especially, we find that the $R_1$ substitution site is able to regulate the HOMO-LUMO gap to a greater extent, which further facilitates the achievement of long-wavelength emission. But is it really what we assume?

To verify the above speculation, the OPA and fluorescence emission properties of the molecules **2a**~**5a** are calculated, and the specific values and simulated spectra are summarized in **Table S7** and **Fig. S10**, respectively. Unsurprisingly, the **4a** and **2a** molecules differ only in the $R_1$ substituent, and the OPA and emission spectra as well as the Stokes shift changes are more significant. This phenomenon cause us to think deeply, if the introduction of appropriate substituents in the $R_1$ site can better achieve our desired results? Therefore, a novel molecular design strategy is proposed, leading to the **2a-n** (**n**=**1-5**) series.

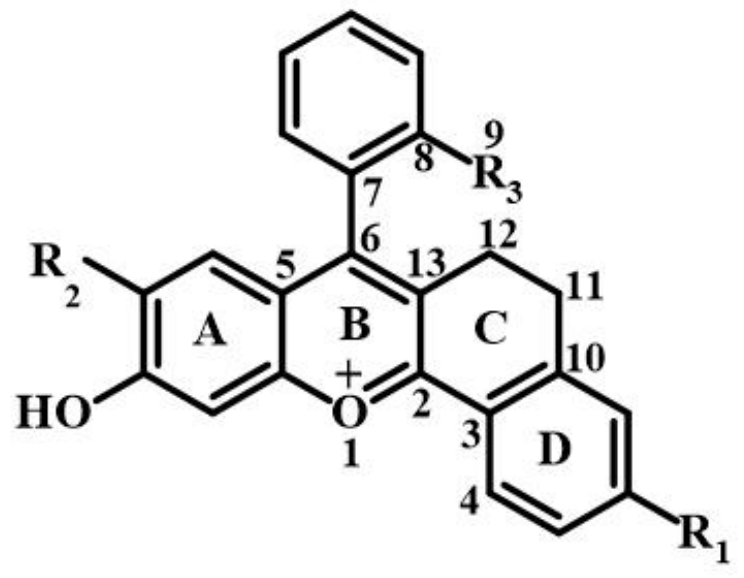

**2a:** $R_1=N(CH_3)_2$ $R_2=Cl$ $R_3=COOH$
**3a:** $R_1=N(CH_3)_2$ $R_2=Cl$ $R_3=OCH_3$
**4a:** $R_1=NH_2$ $R_2=Cl$ $R_3=COOH$
**5a:** $R_1=N(CH_3)_2$ $R_2=H$ $R_3=COOH$

**Fig. S7** The chemical structures of the molecules **2a**~**5a**.

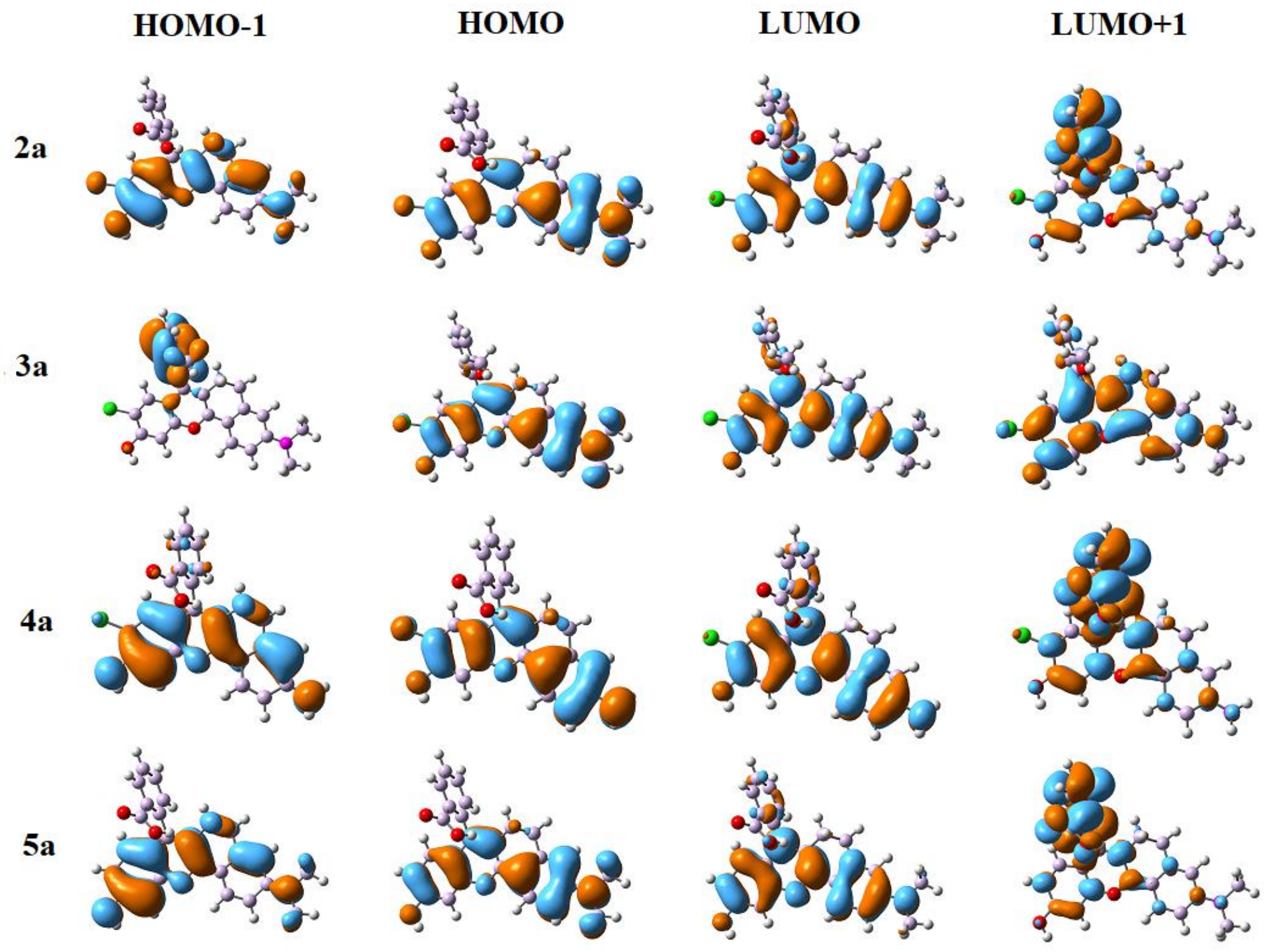

**Fig. S8** The main frontier molecular orbitals of the molecules **2a**~**5a**.

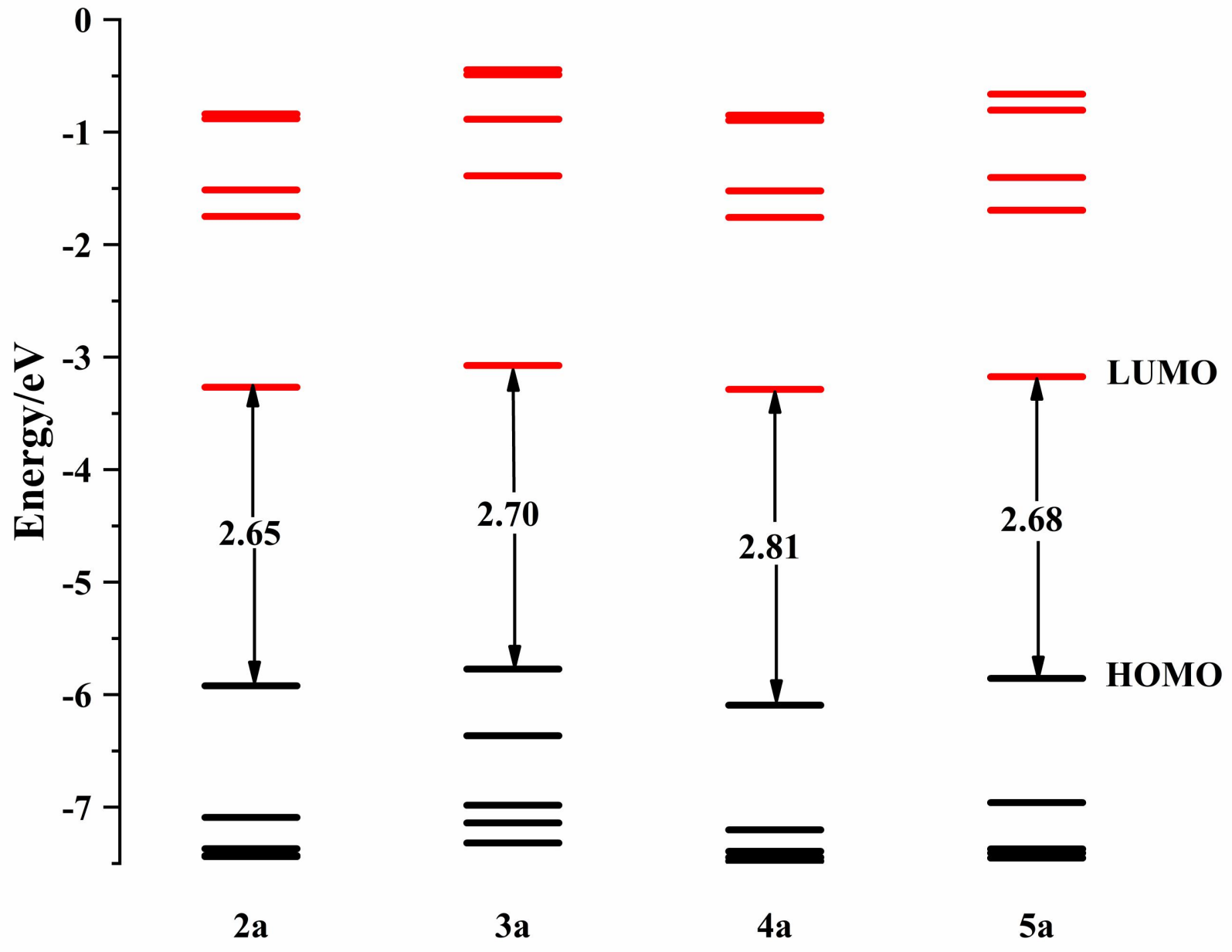

**Fig. S9** The FMO energies of the studied complexes (**2a**~**5a**) by DFT//B3LYP/6-31G(d, p).

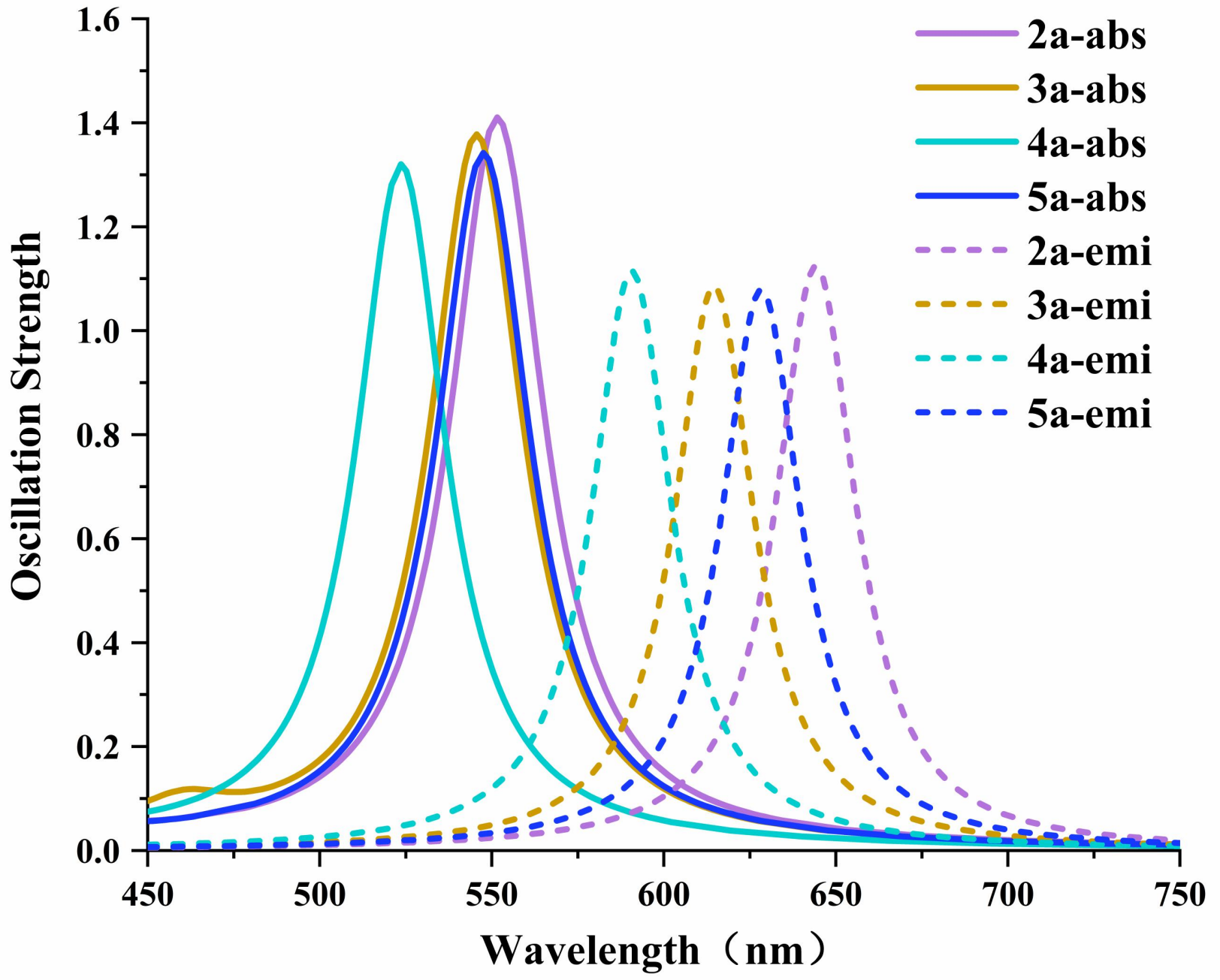

**Fig. S10** The simulated one-photon absorption and fluorescence emission spectra of the studied molecules (**2a**~**5a**).

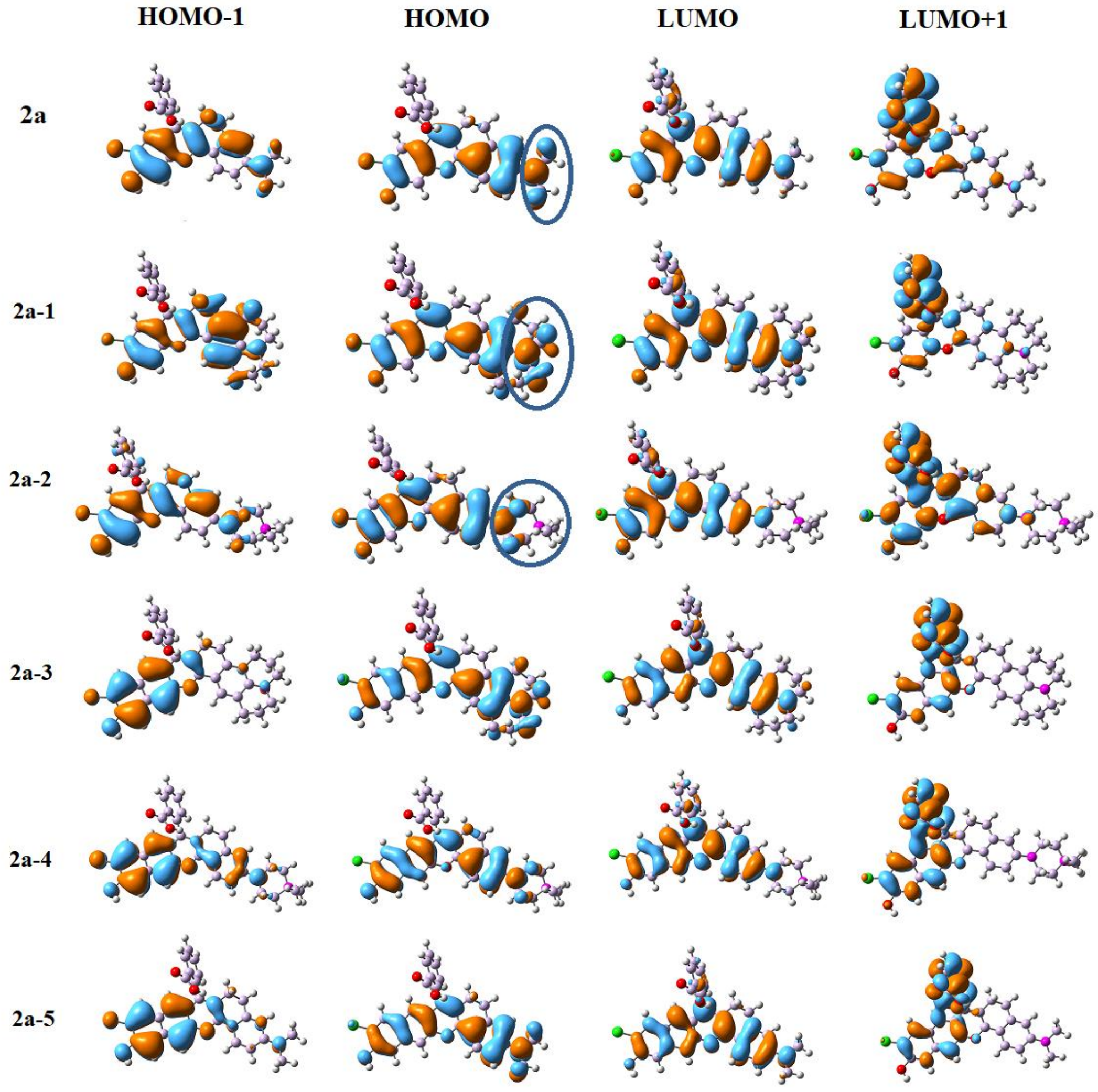

**Fig. S11** The main frontier molecular orbitals of the designed molecules.

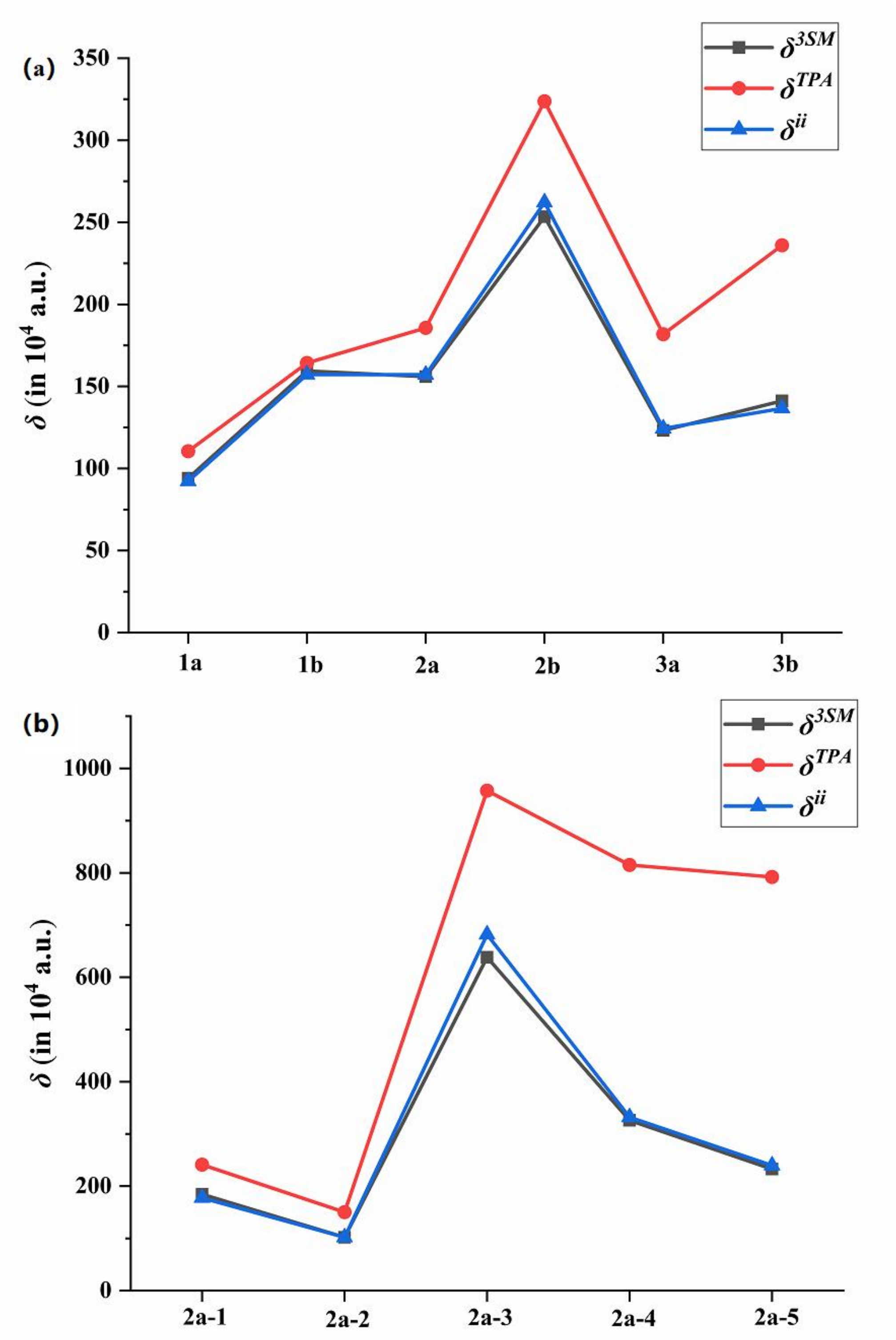

**Fig. S12** Comparison between the $\delta^{3SM}$ calculated by using the three-state model and the $\delta^{TPA}$ predicted by the response theory of the studied molecules. Besides, the $\delta^{ii}$ is the first item in the three-state model formula. (a)the experimental molecules. (b)the designed molecules.

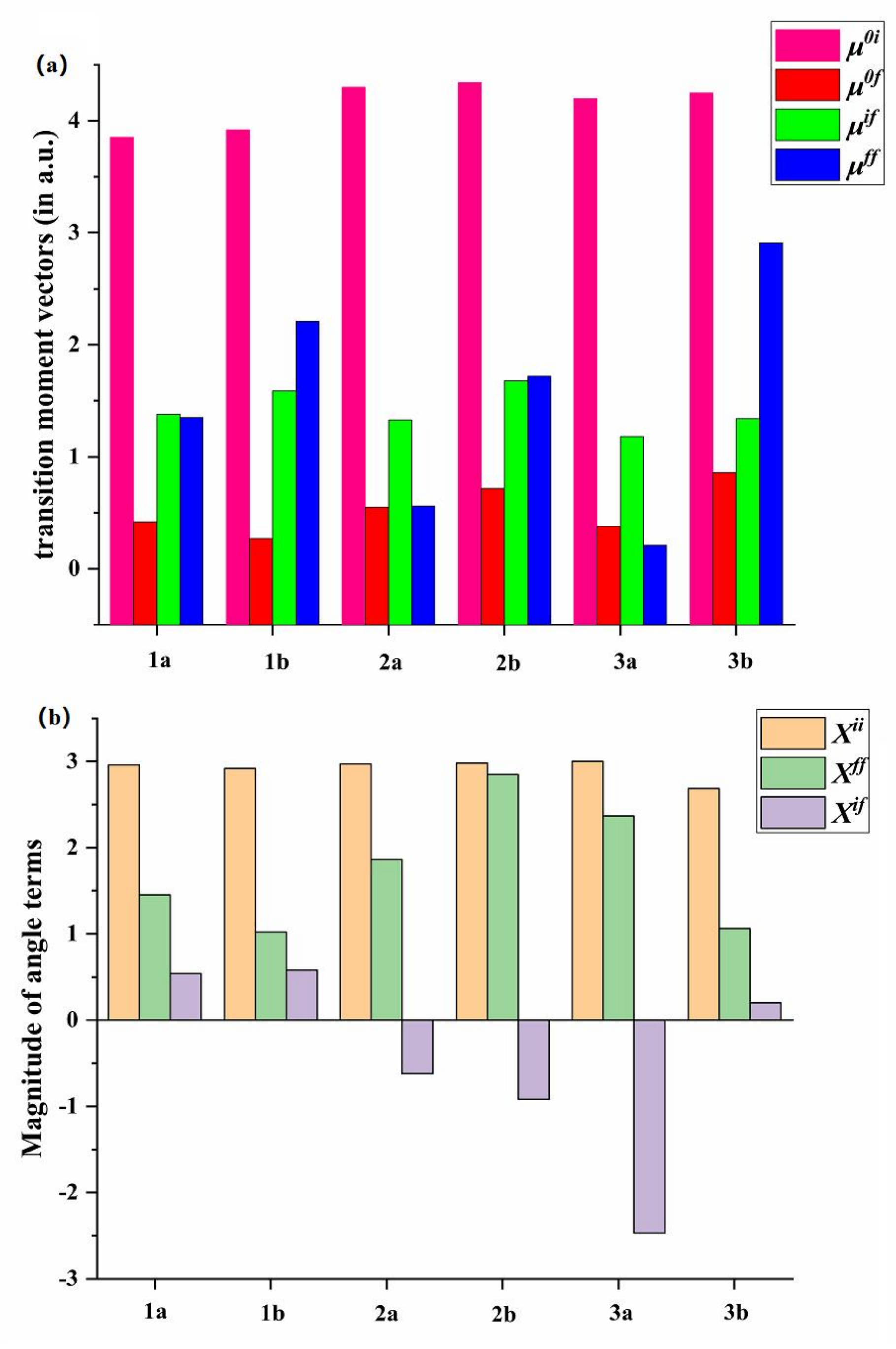

**Fig. S13** Plots of (a) the dipole moment vectors $\mu$ and (b) the angle terms $X$ existed in the 3SM of the studied experimental molecules.

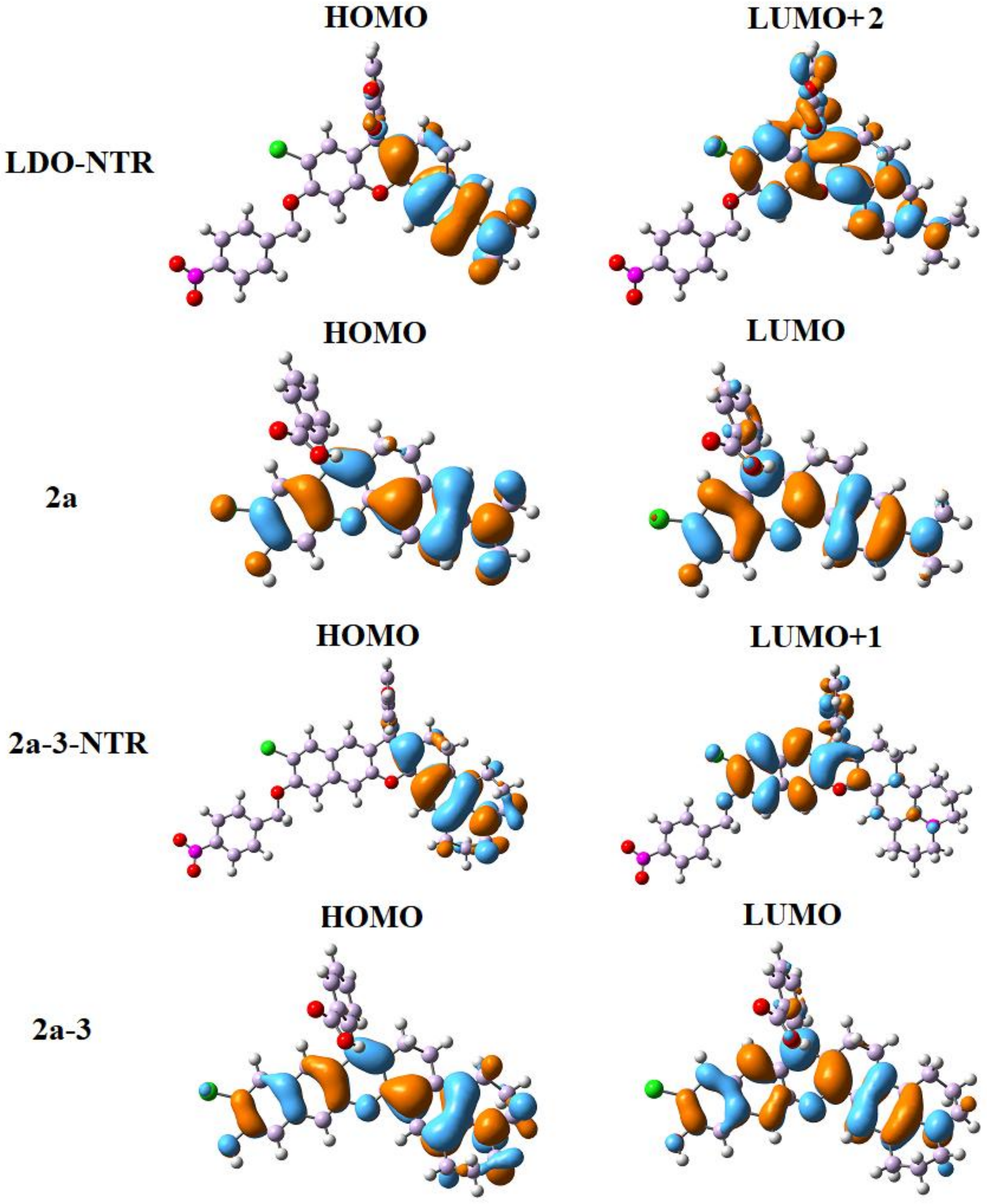

**Fig. S14** The main frontier molecular orbitals of the studied probes and products.

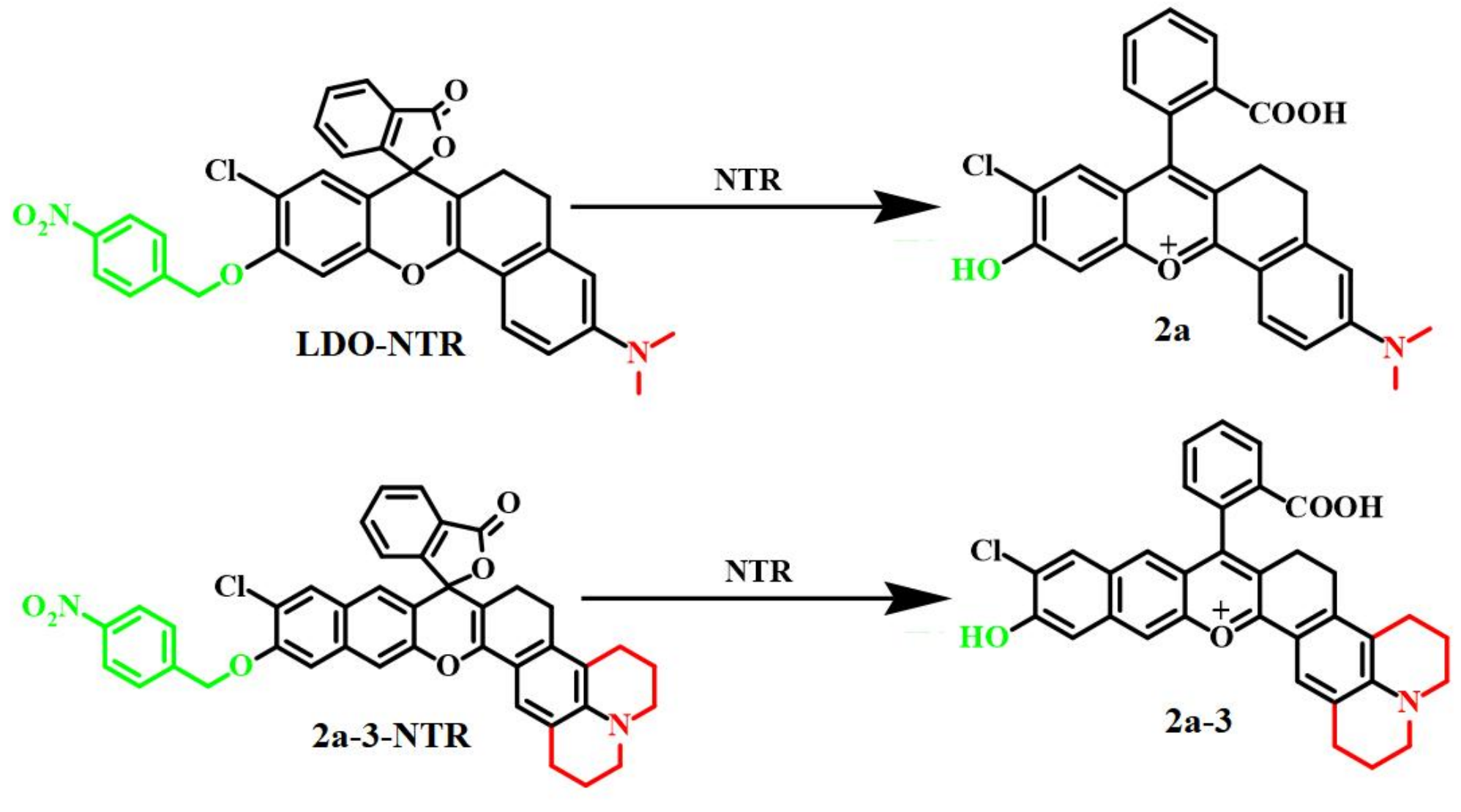

**Fig. S15** Detailed division of the probes and products.(I in green, II in black, III in red)

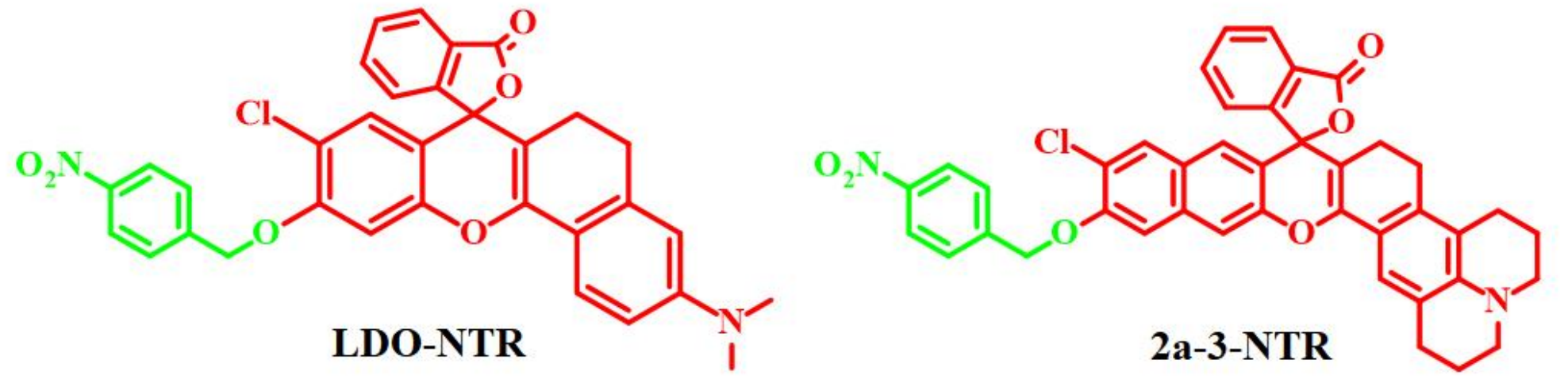

**Fig. S16** The detailed division of electron donor and electron acceptor regions for the probe **LDO-NTR** and **2a-3-NTR**, including the green is the electron acceptor and the red is the electron donor.